\documentclass[journal=jpccck,manuscript=article]{achemso}

\usepackage{chemformula} % Formula subscripts using \ch{}
\usepackage[T1]{fontenc} % Use modern font encodings
\usepackage{amssymb}

\author{Yefei Guo}
\affiliation{Department of Physics, Shanghai University, 99 Shangda Road, 200444 Shanghai, P. R. China}

\author{Jiali Yang}
\affiliation{Department of Physics, Shanghai University, 99 Shangda Road, 200444 Shanghai, P. R. China}

\author{Junhao Zhou} 
\affiliation{Department of Physics, Shanghai University, 99 Shangda Road, 200444 Shanghai, P. R. China}

\author{Na Zhu} 
\affiliation{Department of Physics, Shanghai University, 99 Shangda Road, 200444 Shanghai, P. R. China}

\author{Yichen Jin} 
\affiliation{Department of Physics, Humboldt-Universität zu Berlin, 12489 Berlin, Germany}

\author{G\"unther Thiele}
\affiliation{Institut f\"ur Chemie und Biochemie, Freie Universit\"at Berlin, 14195 Berlin, Germany}

\author{Alexei Preobrajenski}
\affiliation{MAX IV Laboratory, Lund University, 22100 Lund, Sweden}

\author{Elena Voloshina}
\email{elena.voloshina@icloud.com}
\affiliation{Department of Physics, Shanghai University, 99 Shangda Road, 200444 Shanghai, P. R. China}
\alsoaffiliation{Institut f\"ur Chemie und Biochemie, Freie Universit\"at Berlin, 14195 Berlin, Germany}

\author{Yuriy Dedkov}
\email{yuriy.dedkov@icloud.com}
\affiliation{Department of Physics, Shanghai University, 99 Shangda Road, 200444 Shanghai, P. R. China}

\title{Electronic Correlations in Multiferroic van der Waals CuCrP$_2$S$_6$: Insights From X-Ray Spectroscopy and DFT} %Title of paper

\begin{document}


\begin{tocentry}

\includegraphics[width=\textwidth]{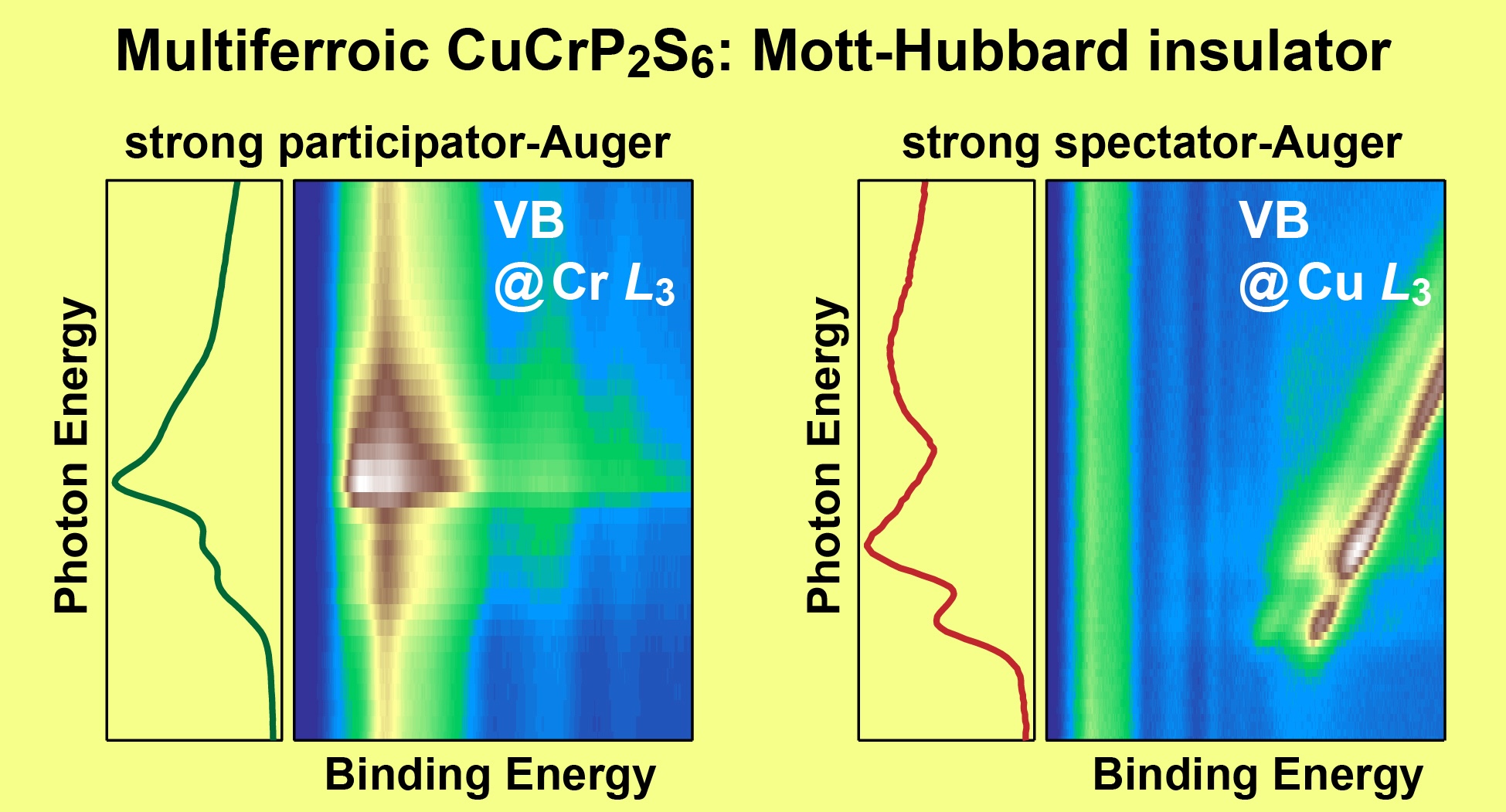}

\end{tocentry}


\begin{abstract}

The electronic structure of high-quality van der Waals multiferroic CuCrP$_2$S$_6$ crystals was investigated applying photoelectron spectroscopy methods in combination with DFT analysis. Using X-ray photoelectron and near-edge X-ray absorption fine structure (NEXAFS) spectroscopy at the Cu\,$L_{2,3}$ and Cr\,$L_{2,3}$ absorption edges we determine the charge states of ions in the studied compound. Analyzing the systematic NEXAFS and resonant photoelectron spectroscopy data at the Cu/Cr\,$L_{2,3}$ absorption edges allowed us to assign the CuCrP$_2$S$_6$ material to a Mott-Hubbard type insulator and identify different Auger-decay channels (participator vs. spectator) during absorption and autoionization processes. Spectroscopic and theoretical data obtained for CuCrP$_2$S$_6$ are very important for the detailed understanding of the electronic structure and electron-correlations phenomena in different layered materials, that will drive their further applications in different areas, like electronics, spintronics, sensing, and catalysis.

\end{abstract}


\section{Intorduction}

The exceptional transport characteristics exhibited by two-dimensional (2D) materials~\cite{Novoselov.2005,Zhang.2005,Manzeli.2017} and recently discovered fascinating properties of different heterosystems on their basis~\cite{Geim.2013,Yu.2014vu,Liu.2016xab,Yankowitz.2019,Pham.2022} have garnered substantial interest to their corresponding three-dimensional (3D) counterparts. These 3D objects are known as layered van der Waals (vdW) materials consisting of individual 2D layers held together by weak vdW bonds. Among them, specifically transition metal phosphorus trichalcogenides (MPX$_3$, M: transition metal, X: chalcogen) reflect their immense potential and prospects across diverse domains~\cite{Samal.2020,Zhang.2023xyz}. These materials exhibit extraordinary properties, including exceptional charge-discharge performance, different magnetic orderings, and a wide range of band gaps of $1.2-3.5$\,eV~\cite{Du.2016,Dedkov.2023ues}. Only recently the electronic structure of MPX$_3$ materials became the subject of systematic studies. For example, near-edge X-ray absorption fine structure (NEXAFS) and resonant photoelectron spectroscopy (ResPES) allowed to assign MnPX$_3$ and FePX$_3$ to the class of Mott-Hubbard insulators~\cite{Kamata.1997,Yang.2020uc,Jin.2022}, NiPS$_3$ to the class of charge-transfer insulators~\cite{Kim.2018,Yan.2021ni}, whereas CoPS$_3$ has a mixed character of the insulating state~\cite{Jin.2022jc}. It is interesting that mixed Fe$_x$Ni$_y$PS$_3$ alloys demonstrate a dual character of the insulating state in their electronic structure pointing to very weak (if any) hybridization of Fe- and Ni-derived electronic states~\cite{Li.2023,Yan.202396s}. Recent angle-resolved photoelectron spectroscopy experiments on different MPX$_3$ materials demonstrate a rather good agreement with available calculated band structures allowing to identify partial elemental and orbital contributions in the electronic structure~\cite{Dedkov.2023ues,Voloshina.2023,Koitzsch.2023,Nitschke.2023,Yan.202396s,Strasdas.2023,Klaproth.2023}.

While a significant amount of research has been dedicated to elucidating the (opto)electronic, magnetic and catalytic properties of different MPX$_3$ materials and alloys~\cite{Du.2018,Wang.2018flk,Dedkov.2023ues}, one particular member of this class, namely mixed CuCrP$_2$S$_6$, remains elusive, with limited reports available on the studies of its electronic and magnetic properties~\cite{Susner.2020,Selter.2023,Wang.2023kp}. Despite its enigmatic nature, CuCrP$_2$S$_6$ holds immense potential for investigation. This material stands out as a captivating member within the realm of MPX$_3$-based 2D multiferroic materials, where antiferroelectric (AFE) and antiferromagnetic (AFM) lattices coexist, each occupying distinct cation sites~\cite{Susner.2020}. Notably, CuCrP$_2$S$_6$ showcases an intriguing arrangement of Cu and Cr on a honeycomb sublattice (Fig.~\ref{fig:CuCrP2S6_Structure_DOS}(a,b)), endowing it with the distinct advantage of tuneable ionic and spin characteristics, making it highly appealing for magnetoelectric storage devices~\cite{Park.2022}.

Recent theoretical studies~\cite{Lai.2019,Park.2022} demonstrate that CuCrP$_2$S$_6$ single layer exhibits intralayer ferromagnetic (FM) ordering of Cr$^{3+}$ ions with a magnetic moment of $\approx3\,\mu_B$ arranged \textit{in plane} and Curie temperature of $T_C=64$\,K. At the same time, neighbouring layers are antiferromagnetically ordered with a Ne\'el temperature of $T_N=31$\,K. The AFE ordering of Cu$^{1+}$ ions is more energetically favourable, giving the opportunity to tune this state via external stimulus. The respective density functional theory (DFT) calculations show a band gap of $0.975$\,eV for the FM/AFE state of CuCrP$_2$S$_6$~\cite{Park.2022}. However, despite the recent progress in the applications-oriented experimental studies of the magnetic, transport and multiferroic properties of CuCrP$_2$S$_6$, there remains a significant lack of the works presenting systematic characterization as well as spectroscopic studies of this material, which provide the most direct insight in the understanding of its electronic and magnetic properties. The present work aims to fill this gap in information and address the most fundamental understanding of electronic and magnetic properties of CuCrP$_2$S$_6$.

Here, we present electronic structure studies of multiferroic vdW CuCrP$_2$S$_6$ using different spectroscopic techniques. X-ray photoelectron spectroscopy (XPS) and NEXAFS confirm the valence states of metal ions as Cu$^{1+}$ and Cr\,$^{3+}$, alongside the charge state of the bi-pyramid [P$_2$S$_6$]$^{4-}$, ensuring the unit cell's charge neutrality. Our systematic ResPES experiments performed at the Cr\,$L_{2,3}$ absorption edge allow us to assign the CuCrP$_2$S$_6$ wide band-gap material to the class of Mott-Hubbard insulators according to the  Zaanen-Sawatzky-Allen scheme. On the other hand, ResPES measurements at the Cu\,$L_{3}$ absorption edge reveal a pronounced feature at large binding energies, away from the main and satellite structures. This structure is attributed to the Cu\,$3d^84s^{1*}$ spectator-Auger decay channel. The obtained spectroscopic results offer valuable insights into the electronic structure and electron-correlation effects in the valence band of CuCrP$_2$S$_6$, that is important for the correct description of the electronic, optical and magnetic properties of this mutiferroic material and for understanding its functionality in different applications. %Details describing experimental and theoretical approaches used in the present study as well as additional data are presented in the Supplementary Material.

\section{Experimental and Theoretical Methods}

\paragraph{Synthesis.} CuCrP$_2$S$_6$ crystals were synthesized using the CVT method outlined in Ref.~\citenum{Selter.2023} (see Fig.\,S1 of Supplementary Material). A quartz ampule used in the synthesis has an inner diameter of $10$\,mm and a wall thickness of $2$\,mm. It was meticulously cleaned by rinsing sequentially with distilled water and isopropanol for $15$ minutes each. Subsequently, the ampule was subjected to a high-temperature treatment in a tube furnace, being baked at $800^{\circ}$\,C for $12$ hours. To initiate the synthesis, a reaction mixture consisting of $0.5$\,g of the ingredients in the molar ratio of Cu : Cr : P : S = 1 : 1 : 2 : 6 was carefully loaded into the prepared and purged quartz ampule. Additionally, $50$\,mg of iodine was included as a transport agent. At the next step, the loaded ampule was evacuated to a vacuum level of $<10^{-4}$\,mbar and then sealed. After that, the sealed ampule was placed horizontally in a two-zone tube furnace. The reactant side (``HOT'') was heated to $750^{\circ}$\,C for a total duration of 274.5 hours, whereas the ``COLD'' side was initially heated to $800^{\circ}$\,C for $24$ hours. Subsequently, the ``COLD'' side was gradually cooled to $700^{\circ}$\,C, facilitating a controlled nucleation. The temperature was then maintained at $700^{\circ}$\,C for $\approx100$ hours to promote a controlled crystal growth. Finally, the ``HOT'' side was rapidly cooled to the ``COLD'' temperature within $1$ hour, followed by gradual cooling of both sides to room temperature. It is worth to note that the used CVT procedure has a yield of approximately $30\%$ for CuCrP$_2$S$_6$ crystals; the second large fraction is the CrPS$_4$ vdW material, which is an AFM semiconductor with a band gap of $\approx1.4$\,eV and $T_N=36$\,K~\cite{Pei.2016,Lee.2017f9o,Susilo.2020} and has very similar XRD patterns~\cite{Lee.2017f9o}. Therefore, a comprehensive sample characterization by several experimental methods has to be applied in order to unequivocally discriminate between these two compounds (see below). 

\paragraph{Characterization.} For the Raman characterization, a Renishaw inVia Qontor was used to acquire high-resolution spectra. CuCrP$_2$S$_6$ crystals were illuminated with a laser of a wavelength of $532$\,nm and a power of $100$\,mW with a $1\,\mu\mathrm{m}$ laser spot. Before characterization, a standard monocrystalline Si wafer was used to calibrate the system. XRD patterns were collected at room temperature with a Desktop Bruker D2 Phaser diffractometer using Cu $K\alpha$ ($\lambda=1.54178$\,\AA) radiation. SEM/EDX data were collected using the ZEISS SIGMA 500 microscope. HR-TEM measurements were performed using an FEI Talos F200x G2 instrument with EDX (super-X) and the FIB preparation was carried out using an FEI Scios 2 HiVac. For the structure refinement experiments CuCrP$_2$S$_6$ single crystals were isolated and collected under an optical microscope, mounted in Paraton$^\circledR$ oil, and measured using a Bruker D8 Venture diffractometer with Mo\,$K\alpha$ radiation ($\lambda = 0.71073$\,\AA) at variable temperature. The crystal structures were processed and refined in Olex\,2~\cite{Dolomanov.2009} using ShelXT~\cite{Sheldrick.2015} and ShelXL~\cite{Sheldrick.2015kit} programs.

\paragraph{XPS, NEXAFS and ResPES.} Initially, CuCrP$_2$S$_6$ crystals were studied in the laboratory-based XPS UHV station installed at Shanghai University and consisting of preparation and analysis chambers with a base pressure better than $1\times10^{-10}$\,mbar (SPECS Surface Nano Analysis GmbH). XPS spectra were collected using a monochromatized Al\,$K\alpha$ ($h\nu = 1486.6$\,eV) X-ray source and SPECS PHOIBOS 150 hemispherical analyzer combined with a 2D-CMOS detector. These data are presented below. NEXAFS and ResPES experiments were performed at the EA01 endstation for the PES and NEXAFS experiments of the SMS branch of the FlexPES beamline (MAX\,IV synchrotron radiation facility, Lund, Sweden)~\cite{Preobrajenski.2023}. All spectra were measured in UHV conditions (base vacuum is below $1 \times 10^{-10}$\,mbar) and at room temperature. NEXAFS spectra were collected in the partial electron yield (PEY) mode using a channel-plate detector with a grid repulsive voltage of $U = -400$\,V and $U = -500$\,V for Cr\,$L_{2,3}$ and Cu\,$L_{2,3}$ absorption edges, respectively. XPS core-level and PES valence band spectra were acquired using a ScientaOmicron DA\,$30$-L(W) energy analyzer. ResPES spectra were collected in the fixed mode of the energy analyzer, while the photon energy was scanned across the Cr\,$L_{2,3}$ and Cu\,$L_{3}$ absorption edges with a step of $0.2$\,eV. The obtained valence band PES intensity maps $I(E_B, h\nu)$ were then used for the analysis ($E_B$ is the binding energy; $h\nu$ is the photon energy). For all spectroscopic experiments CuCrP$_2$S$_6$ crystals were mounted on the Mo sample holder using Ta-foil stripes. In order to obtain clean surfaces, crystals were cleaved in air under flow of dry N$_2$ gas using scotch tape and then introduced in vacuum within the next $30$\,sec. The cleanliness of samples was verified using XPS of core levels and valence band. Before every set of XPS experiments, the freshly cleaned Ag-poly or Au-poly sample was used for the calibration of the Fermi level of the instrument. Additionally, freshly cleaned Cu-poly sample was measured for the reference core level and valence band XPS spectra.

 \paragraph{Theory.}  The spin-polarized DFT calculations were carried out with the Vienna \textit{ab initio} simulation package (VASP)~\cite{Kresse.1996uh,Kresse.1999}, employing the generalised gradient approximation (GGA) functional of Perdew, Burke and Ernzerhof (PBE)~\cite{Perdew.1996}. The ion-cores were described by projector-augmented-wave (PAW) potentials~\cite{Blochl.1994} and the valence electrons [Cr ($3d$, $4s$), Cu ($3d$, $4s$), P ($3s$, $3p$), and S ($3s$, $3p$)] were described by plane waves associated to kinetic energies of up to $400$\,eV. Brillouin-zone integration was performed on $\Gamma$-centred symmetry with Monkhorst-Pack meshes by Gaussian smearing with $\sigma = 0.05$\,eV except for density of states (DOS) calculations. For DOS calculations, the tetrahedron method with Bl\"ochl corrections was employed~\cite{Blochl.1994abc}. Since our calculations were performed for a hexagonal unit cell (see Fig.\,S2 of Supplementary Material), the $12 \times 12 \times 2$ $k$-mesh was used. The convergence criteria for energy was set equal to $10^{-5}$\,eV. The van der Waals interactions were incorporated by the semi-empirical approach of Grimme through the D2 correction~\cite{Grimme.2006}. The lattice parameters of 3D CuCrP$_2$S$_6$ (lattice vectors and positions of Cu, P, and S) were fully relaxed. During structure optimization, the convergence criteria for force was set equal to $10^{-2}$\,eV\,\AA$^{-1}$. The DFT$+U$ scheme~\cite{Anisimov.1997,Dudarev.1998} was adopted for the treatment of Cr\,$3d$ and Cu\,$3d$ orbitals, with the parameter $U_\mathrm{eff} = U - J$ equal to $4$\,eV. This way, the optimised lattice parameters (see Table\,T1 of Supplementary Material), the calculated distribution of electronic states in the density of states, valencies of elements, and the band gap (see below) are in good agreement with experimental data.  For comparison reasons, the DOS calculations were performed using the HSE06 functional~\cite{Heyd.2005} as well as another combination of $U_\mathrm{eff}$ ($U_\mathrm{eff}=5$\,eV~\cite{Rohrbach.2004} and $7$\,eV~\cite{Nolan.2006}, for Cr\,$3d$ and Cu\,$3d$ orbitals respectively). The results are presented in Figure\,S3 of Supplementary Material.

\section{Results and Discussion}

\begin{figure}
\includegraphics[width=\textwidth]{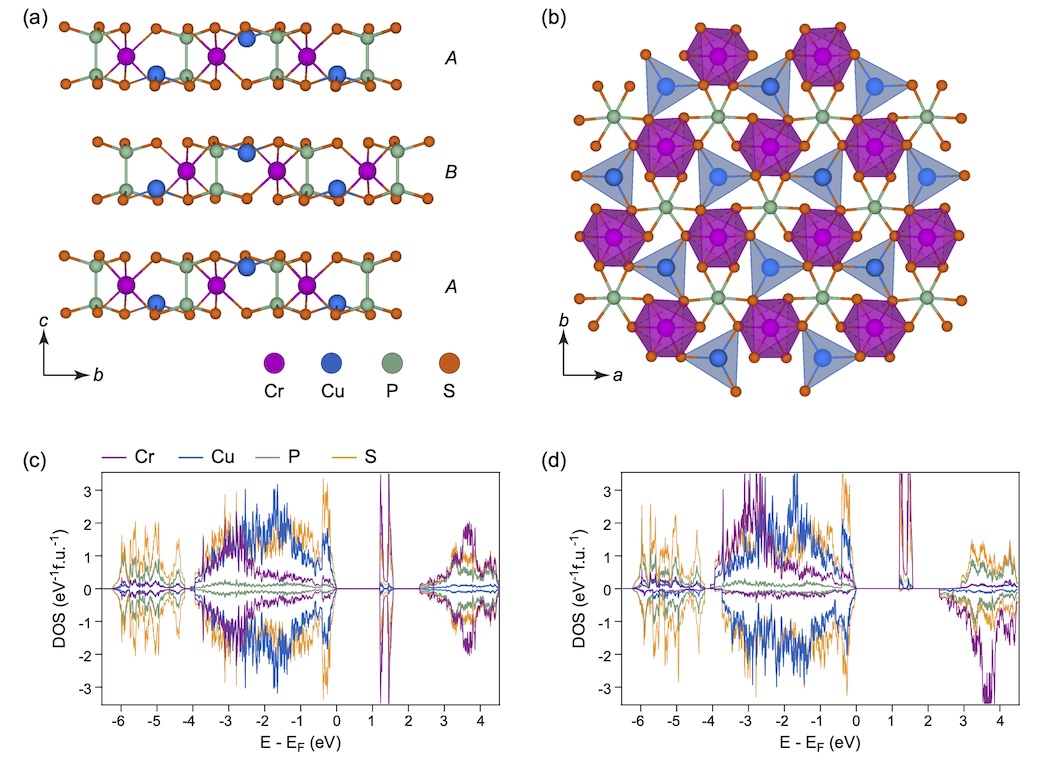}
\caption{\label{fig:CuCrP2S6_Structure_DOS} (a,b) Crystallographic structures of bulk (side view) and single layer (top view) of CuCrP$_2$S$_6$. (c,d) Corresponding DOS plots for the ground state AFM bulk and FM monolayer CuCrP$_2$S$_6$.}
\end{figure}

Bulk CuCrP$_2$S$_6$ crystallizes in the $C2/c$ space group and contains $AB$-stacked 2D layers bonded via the vdW interactions (Fig.~\ref{fig:CuCrP2S6_Structure_DOS}(a)). Each of these 2D layers is made of quasi-trigonal CuS$_3$, octahedral CrS$_6$, and P$_2$S$_6$ units, where Cu and Cr ions form a honeycomb lattice in an ordered way in the $ab$-plane (Fig.~\ref{fig:CuCrP2S6_Structure_DOS}(b)). The Cr ions are centered in every 2D layer, whereas the Cu ions are strongly displaced either towards the bottom or top layer of sulphur. This structure can also be represented in hexagonal unit cell containing $6$ layers (used in this work; see Fig.\,S2 of Supplementary Material for representations). 
Our calculations using the hexagonal unit cell yield $a=11.903$\,\AA\ and $c=39.12$\,\AA. The obtained in DFT calculations lattice parameters of CuCrP$_2$S$_6$ corresponding to the $C2/c$ space group are presented in Table\,T1 of Supplementary Material.

According to our DFT calculations, 3D bulk CuCrP$_2$S$_6$ is an AFM semiconductor in its ground state, with a band gap of $1.2$\,eV. The calculated band gap is in a good agreement with a value of $1.239$\,eV obtained in optical absorption studies~\cite{Studenyak.2003}. At that, the Cr ions within one layer are FM coupled to each other, while antiferromagnetically coupled to Cr ions in the adjacent layers. The calculated density of states (DOS) plots for AFM bulk and FM single layer of CuCrP$_2$S$_6$ are presented in Fig.~\ref{fig:CuCrP2S6_Structure_DOS}(c,d), respectively. According to these results, the local magnetic moment of $3.402\,\mu_B$ is concentrated on Cr$^{3+}$ ions, whose electronic states are strongly exchange split. This leads to the appearance of energy-localized states at the bottom of the conduction band, which mainly possess the Cr\,$3d$ character. For AFM 3D bulk CuCrP$_2$S$_6$ the vdW interaction between layers is weak and the total DOS is a combination of spin-polarized contributions from adjacent layers, with a total magnetic moment of zero.

The CuCrP$_2$S$_6$ crystals synthesised in the present work have lateral dimensions of \mbox{$\approx5\times5\,\mathrm{mm}^2$} and thickness of several hundreds $\mu\mathrm{m}$ (see insets of Fig.~\ref{fig:CuCrP2S6_XRD_Raman_SEM_TEM}(a) and Fig.\,S4 of Supplementary Material). The layered structure of crystals can be easily identified in optical microscopy images. Fig.~\ref{fig:CuCrP2S6_XRD_Raman_SEM_TEM}(a) demonstrates representative XRD patterns for the CuCrP$_2$S$_6$ crystal. According to the layered structure and $C2/c$ symmetry group of CuCrP$_2$S$_6$, sharp $(00l)$ reflexes ($l$ is even) are observed confirming the high crystallographic quality of the crystals. The series of XRD patterns collected over $30$ CuCrP$_2$S$_6$ crystals gives the mean value for the $(004)$ spot as $27.655^\circ$ that corresponds to the distance of $6.644$\,\AA\ between CuCrP$_2$S$_6$ layers (see Fig.\,S5 of Supplementary Material). Further results on the structure refinement for the CuCrP$_2$S$_6$ crystal are presented in Supplementary Material and give very good agreement with DFT results. 

The Raman spectrum of CuCrP$_2$S$_6$ measured at room temperature is presented in Fig.~\ref{fig:CuCrP2S6_XRD_Raman_SEM_TEM}(b). A comparison of these data with previously published results demonstrates a very good agreement between them~\cite{Susner.2020,Io.2023,Rao.2023,Ma.2023}. Specifically, in the Raman spectrum of CuCrP$_2$S$_6$ four distinct peaks can be identified, which correspond to: rotation (R(PS$_3$), $201.55\,\mathrm{cm}^{-1}$) and translation (T(PS$_3$), $263.38\,\mathrm{cm}^{-1}$) of the PS$_3$ group, respectively, and out-of-plane P-P ($\nu$(P-P), $375.64\,\mathrm{cm}^{-1}$) and P-S ($\nu$(P-S), $584.61\,\mathrm{cm}^{-1}$) stretching vibrations within the [P$_2$S$_6$]$^{4-}$ ethane-like building blocks of the anion sublattice, respectively~\cite{Susner.2020}. As was previously shown, both, XRD and Raman spectroscopy, are used to study AFE and AFM transitions, which were confirmed in CuCrP$_2$S$_6$~\cite{Susner.2020,Io.2023,Rao.2023}. 

Further analysis of the bulk structure and stoichiometry of the CuCrP$_2$S$_6$ crystals was performed using combined SEM/EDX and TEM/EDX methods (Fig.~\ref{fig:CuCrP2S6_XRD_Raman_SEM_TEM}(c,d) and Fig.\,S7 of Supplementary Material). The atomic concentration analysis using SEM/EDX confirms the correct stoichiometry for the CuCrP$_2$S$_6$ crystal pointing to the successful synthesis of the bulk crystals with the desired elemental composition (Tab.\,T2 of Supplementary Material). The TEM analysis of CuCrP$_2$S$_6$ (Fig.~\ref{fig:CuCrP2S6_XRD_Raman_SEM_TEM}(d)) gives a distance between single planes of $6.475$\,\AA, along with a slight excess of Cu concentration over Cr.

\begin{figure}
\includegraphics[width=\textwidth]{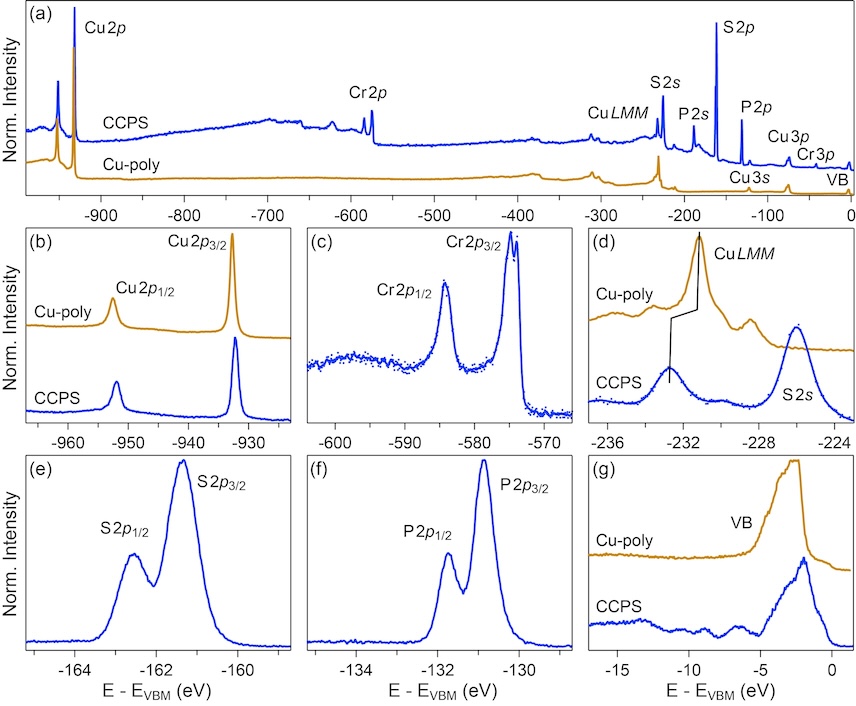}
\caption{\label{fig:CuCrP2S6_XPS_MAXIV_overview} XPS spectra of freshly cleaved CuCrP$_2$S$_6$ (CCPS) bulk crystal and Cu-poly reference sample: (a) survey, (b) Cu\,$2p$, (c) Cr\,$2p$, (d) Cu\,$LMM$ Auger lines, (e) S\,$2p$, (f) P\,$2p$, and (g) valence band. All spectra were collected with photon energy $h\nu=1150$\,eV.}
\end{figure}

Initial XPS experiments on CuCrP$_2$S$_6$ were performed under laboratory conditions as described in the Supplementary Material. The XPS survey spectrum of the freshly cleaved CuCrP$_2$S$_6$ crystal collected at the synchrotron light source exhibits all characteristic emission lines without visible oxygen- and carbon-related peaks confirming the high quality and cleanliness of the studied samples (Fig.~\ref{fig:CuCrP2S6_XPS_MAXIV_overview}(a)). The high-resolution Cu\,$2p$ XPS and Cu\,$L_3M_{45}M_{45}$ Auger spectra confirm the Cu$^{1+}$ charge state~\cite{Wang.2018oja,Liu.2022} (Fig.~\ref{fig:CuCrP2S6_XPS_MAXIV_overview}(b,d)). The obtained energy difference between positions of the Cu\,$L_3M_{45}M_{45}$ Auger peak for CuCrP$_2$S$_6$ and Cu-poly sample is $\approx1.7$\,eV, in a very good agreement with previous results~\cite{Wang.2018oja}. The high-resolution Cr\,$2p$ XPS spectrum of CuCrP$_2$S$_6$ (Fig.~\ref{fig:CuCrP2S6_XPS_MAXIV_overview}(c)) is very similar to the one measured for Cr$_2$O$_3$, where the complex shape of this spectrum is due to the coupling between the $2p$ core-hole and the unpaired electrons in the $3d$ outer shell~\cite{Chambers.2001,Biesinger.2004,Bataillou.2020,Pinho.2021}, thus confirming the Cr$^{3+}$ charge state of chromium ions in the studied vdW crystal. The high-resolution S\,$2p$ and P\,$2p$ XPS spectra of CuCrP$_2$S$_6$ (Fig.~\ref{fig:CuCrP2S6_XPS_MAXIV_overview}(e,f)) demonstrate single spin-orbit split doublets with the energy splitting of $1.2$\,eV and $0.87$\,eV, respectively, indicating the presence of only one chemical state for both S and P ions.

\begin{figure}
\includegraphics[width=0.5\textwidth]{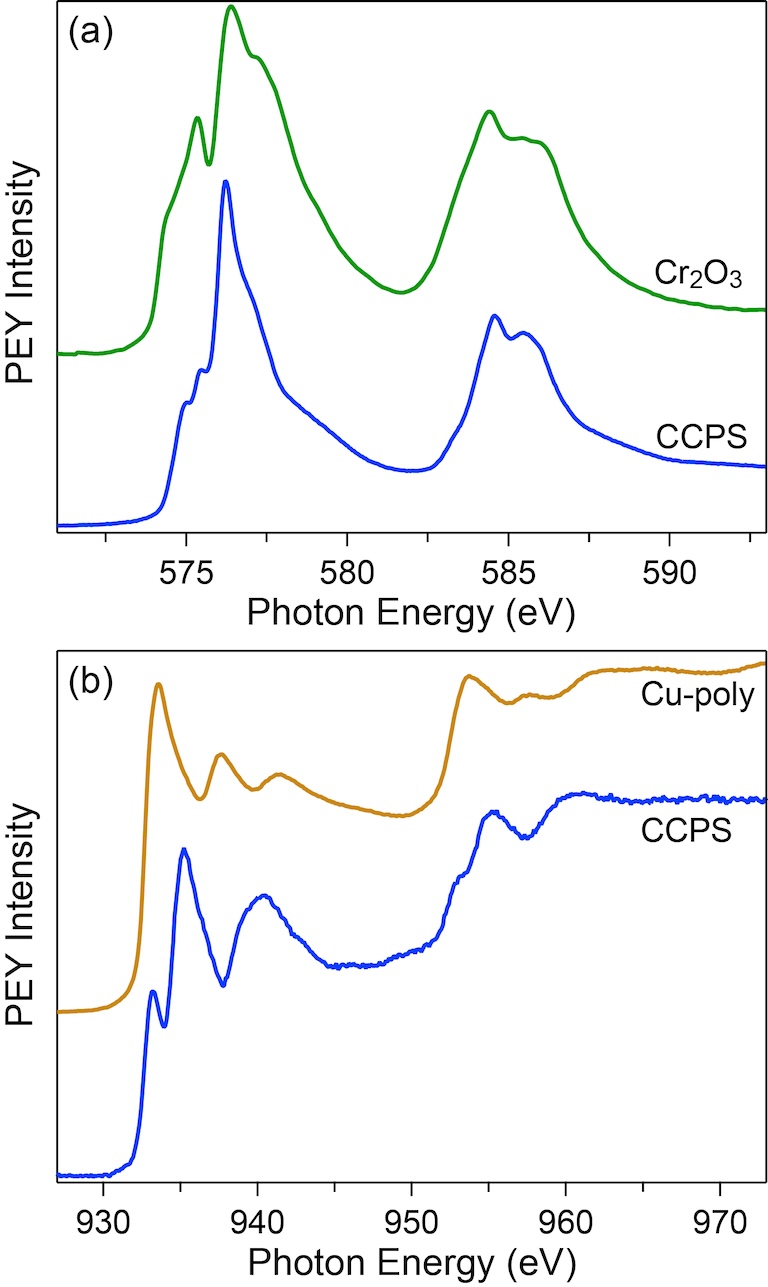}
\caption{\label{fig:CuCrP2S6_NEXAFS_MAXIV} NEXAFS spectra of CuCrP$_2$S$_6$ (CCPS) collected in the PEY mode at the (a) Cr\,$L_{2,3}$ and (b) Cu\,$L_{2,3}$ absorption edges. The respective reference spectra for Cr$_2$O$_3$ (Ref.~\citenum{Dedkov.2005}) and Cu-poly sample (present experiment) are also shown.}
\end{figure}

Figure~\ref{fig:CuCrP2S6_NEXAFS_MAXIV} presents (a) Cr\,$L_{2,3}$ and (b) Cu\,$L_{2,3}$ NEXAFS spectra of CuCrP$_2$S$_6$ together with the respective reference X-ray absorption spectra for Cr$_2$O$_3$ (Cr$^{3+}$) and Cu metal (Cu$^0$). The Cr\,$L_{2,3}$ absorption spectrum of CuCrP$_2$S$_6$ closely resembles that of Cr$_2$O$_3$~\cite{Theil.1998,Dedkov.2005} confirming the Cr$^{3+}$ charge state of octahedrally coordinated chromium ions in the studied vdW material. The observed shape variations of the $L_3$ spectrum for two Cr-based compounds can be attributed to a different octahedral crystal-field strength in CuCrP$_2$S$_6$ (CrS$_6$ octahedra) and Cr$_2$O$_3$ (CrO$_6$ octahedra) as well as different multiplet splitting for the Cr\,$3d$ orbitals in the final state~\cite{Theil.1998,Brik.2004}.

The Cu\,$L_{2,3}$ NEXAFS spectrum of CuCrP$_2$S$_6$ is, to some extent, similar to the ones for other Cu$^{1+}$ compounds, like Cu$_2$O or Cu$_2$S~\cite{Grioni.1991,Vegelius.2012,Jiang.2013,Henzler.2015}, however with the intensity of the first peak suppressed. In Cu metal copper has a formal valence state Cu(0) ($d^{10}s^1$) and the respective threshold structure in the NEXAFS reference spectrum is explained as a result of the residual $3d-4s$ hybridization above the Fermi level~\cite{Leapman.1982,Muller.1982}. Correspondingly, the same explanation is valid for the NEXAFS spectrum of Cu$_2$O with a formal valence state Cu(I) ($d^{10}s^0$), where residual $3d$ partial DOS is obtained in the DFT calculations~\cite{Grioni.1991}. However, in this case the renormalization of the $3d$ partial density of states due to the large Cu\,$2p$ core-hole potential caused by the small extra-atomic screening leads to the sharp peak in the Cu\,$L_{2,3}$ NEXAFS spectrum of Cu$_2$O with a wide band-gap of $\approx2.1-2.6$\,eV~\cite{Heinemann.2013,Visibile.2019}. Comparison with the present results for CuCrP$_2$S$_6$ reveals a suppression of intensity for the peak located at the absorption threshold, indicating more efficient screening for the core-hole, which is possibly due to the electrons which are promoted to states located at $E-E_{VBM}\approx1.3$\,eV (see Fig.~\ref{fig:CuCrP2S6_Structure_DOS}(c,d)).

Further investigations of the electronic structure of CuCrP$_2$S$_6$ and effects of electron-correlations in this material were performed using ResPES of the valence band states at the Cr\,$L_{2,3}$ and Cu\,$L_3$ absorption edges (Fig.~\ref{fig:CuCrP2S6_ResPES_MAXIV}). Fig.\,S10 of Supplementary Material also presents the ``ON'' ($h\nu=576.21$\,eV) and ``OFF'' ($h\nu=570$\,eV) resonance spectra measured in a wide energy range at the Cr\,$L_{3}$ absorption edge. Generally, the valence band XPS spectrum of MPX$_3$ materials, which are wide band-gap semiconducting materials, typically exhibits a wide intensive peak in the vicinity of the Fermi level, which is assigned to the $3d^n\underline{L}$ final state for the charge-transfer type insulators ($U_{dd}>\Delta$) and $3d^{n-1}$ final state for the Mott-Hubbard type insulator ($U_{dd}<\Delta$), and a series of weaker satellites at higher binding energies, which are assigned to the opposite final states in both cases, respectively ($\underline{L}$ is the hole on the ligand site; $U_{dd}$ is the $d-d$ correlation energy; $\Delta$ is the charge transfer energy between $d$-states of the metal and $p$-states of the ligand)~\cite{Zaanen.1985,Yan.2021ni, Kamata.1997,Yang.2020uc,Jin.2022,Jin.2022jc,Li.2023,Yan.202396s,Dedkov.2023ues}. This is also valid for CuCrP$_2$S$_6$, whose valence band XPS spectrum is shown in Fig.~\ref{fig:CuCrP2S6_XPS_MAXIV_overview}(g). In ResPES of MPX$_3$ materials (see schemes in Fig.\,S11 of Supplementary Material) the interference between the processes of (i) direct photoemission ($2p^63d^n + h\nu \rightarrow 2p^63d^{n-1} + e$) and (ii) photoabsorption followed by a participator-Auger Coster-Kronig decay ($2p^63d^n + h\nu \rightarrow 2p^53d^{n+1} \rightarrow 2p^63d^{n-1} + e$) leads to the Fano-type resonance for the identical final state $3d^{n-1}$. This allows us to identify relative energy positions of the respective final states in the spectra ($3d^{n-1}$ and $3d^n\underline{L}$) and determine the type of the insulating states of the MPX$_3$ material. Additionally, the existence of the spectator-Auger decay channel is possible, which spectral bands appear in the spectra at higher binding energies as a result of the final-state screening by the remaining electron on the unoccupied orbitals of the studied material (Fig.\,S11 of Supplementary Material). It is worth to mention, that according to the considered energy diagram presented in Fig.\,S11 of Supplementary Material, the spectral features of the participator-Auger decay have constant binding energies, while spectator-Auger decay spectral features have (nearly) constant kinetic energies, respectively, when photon energy is changed in ResPES studies.

Following this consideration, the resonance behaviour of the valence band states of CuCrP$_2$S$_6$ at the Cr\,$L_{2,3}$ absorption edge can be assigned to the interference process between direct photoemission channel and the participator-Auger process (Fig.~\ref{fig:CuCrP2S6_ResPES_MAXIV}(a)), whereas it is strongly dominated by the spectator-Auger decay at the Cu\,$L_{3}$ edge (Fig.~\ref{fig:CuCrP2S6_ResPES_MAXIV}(b)). Indeed, in case of ResPES at Cr\,$L_{2,3}$ a strong enhancement of the main photoemission band in the energy range up to $E-E_{VBM}\approx-5$\,eV is observed. Given the fact that intensity of the $3d^{n-1}$ (Cr\,$3d^2$) final state is strongly enhanced, we conclude that CuCrP$_2$S$_6$ can be assigned to the class of Mott-Hubbard insulators, similar to MnPX$_3$ and FePX$_3$ materials~\cite{Zaanen.1985,Yang.2020uc,Jin.2022}. The respective binding energy position of the resonating photoemission band remains the same during the photon energy scanning across the Cr\,$L_{2,3}$ absorption edge confirming the strong contribution of the participator-Auger process in the resonance. The corresponding spectator-Auger decay emission can be observed as a weak photoemission intensity above the Cr\,$L_{2,3}$ absorption edges, whose binding energy scales linearly with the photon energy (constant kinetic energy).

In case of ResPES at the Cu\,$L_{3}$ absorption edge (Fig.~\ref{fig:CuCrP2S6_ResPES_MAXIV}(b)) there is no intensity increase for the main emission band just below the valence band maximum ($E-E_{VBM}>-5$\,eV) and the intensity of the emission satellites in the energy range of $E-E_{VBM}\approx-5\cdots-25$\,eV is increased by only approx. $20$\%. As it was discussed before, the formal valence state of copper ions in CuCrP$_2$S$_6$ is Cu$^{1+}$ ($3d^{10}4s^0$). Therefore, photoabsorption at the Cu\,$L_{2,3}$ edge leads to the intermediate state with the formal occupancy $3d^{10}4s^{1*}$. Similar to the previously discussed case of Cu$_2$O, the residual $3d-4s$ hybridization might lead to the appearance of the Cu\,$3d$ character for the unoccupied states; however, its density is small. Therefore, due to the small Cu\,$3d$ density of unoccupied states and insignificance of the $2p\rightarrow4s$ transitions in the ResPES processes, the participator-Auger decay channel is strongly suppressed, preventing any resonant enhancement for the Cu\,$3d$ valence band states. Therefore, the emission at the top of the valence band has a contribution of only Cu\,$3d^9$ final state.

In contrast, several strong emission bands at the energy of $E-E_{VBM}<-11$\,eV, with the intensity increase by factor of more than 14 compared to the background intensity taken for the pre-edge spectrum, are observed at photon energies corresponding to the Cu\,$L_3$ absorption edge (Fig.~\ref{fig:CuCrP2S6_ResPES_MAXIV}(b)). These bands have constant kinetic energies and in the presented plot their binding energies shift linearly as a function of photon energy. Therefore, based on the scheme of the possible ResPES decay channels (Fig.\,S11 of Supplementary Material) and the observation that these bands are located at higher binding energies than expected, we can infer the spectator-Auger decay character of these emission lines. This corresponds to the $3d^84s^{1*}$ final state with the excited electron localized in the conduction band of CuCrP$_2$S$_6$. This assignment is further confirmed by the similarity in spectral shapes for the spectator-Auger signal and the normal Cu\,$L_3M_{45}M_{45}$ Auger signal collected with photon energies away from the Cu\,$L_3$ absorption edge (Fig.~\ref{fig:CuCrP2S6_XPS_MAXIV_overview}(d)).

\section{Conclusions}

In summary, we have synthesised high-quality multiferroic CuCrP$_2$S$_6$ vdW crystals, and characterized them using different methods, including XRD, Raman spectroscopy, SEM/TEM combined with EDX spectroscopy. Further XPS and NEXAFS studies confirm the formal valences of metal ions as Cu$^{1+}$ and Cr$^{3+}$, aligning closely with the occupancy numbers obtained from the calculated ground state electronic structure, being a foundation for the understanding of electronic properties and correlation effects in CuCrP$_2$S$_6$. These effects were addressed in a series of NEXAFS and ResPES experiments performed at the Cu/Cr\,$L_{2,3}$ absorption edges. Results obtained at the Cr\,$L_{2,3}$ absorption edge indicate a strong contribution of the participator-Auger decay in the resonance process, with the Cr\,$3d^{n-1}$ final state identified at the top of the valence band. This allows us to assign CuCrP$_2$S$_6$ to the Mott-Hubbard class of insulators according to the  Zaanen-Sawatzky-Allen scheme. Similar measurements performed at the Cu\,$L_{3}$ edge reveal the Cu\,$3d^84s^{1*}$ final state, as manifested in a strong spectator-Auger decay. All spectroscopic results correlate well with the obtained theoretical data, providing valuable insights into the electronic structure and electron-correlation effects in CuCrP$_2$S$_6$. The behaviour of multiferroic CuCrP$_2$S$_6$ material in different recently proposed applications is based on the interplay between different degrees of freedom, like charge, spin, and valley number. Therefore the information obtained in the present study is crucial for accurately describing and understanding CuCrP$_2$S$_6$ and other layered wide band-gap materials, that will help to adopt them for various future applications in (opto)spintronics, sensing and catalysis.

 \begin{acknowledgement}

The authors thank the National Natural Science Foundation of China (Grant No. 21973059) for financial support. The authors gratefully acknowledge the computing time made available to them on the high-performance computer ``Lise'' at the NHR Center NHR@ZIB. This center is jointly supported by the Federal Ministry of Education and Research and the state governments participating in the NHR (www.nhr-verein.de/unsere-partner). We acknowledge MAX\,IV Laboratory for time on Beamline FlexPES under Proposals 20230147 and 20230148. Research conducted at MAX IV, a Swedish national user facility, is supported by the Swedish Research council under contract 2018-07152, the Swedish Governmental Agency for Innovation Systems under contract 2018-04969, and Formas under contract 2019-02496. Jian Yuan and Yanfeng Guo are acknowledged for the single-crystal XRD measurements.

\end{acknowledgement}

\begin{suppinfo}

The following files are available free of charge.
\begin{itemize}
  
\item Additional experimental and theoretical data (PDF).
%Supplementary description (Description of experimental and theoretical methods; Structure refinement for CuCrP$_2$S$_6$; Laboratory-based XPS characterization of CuCrP$_2$S$_6$; XRD, Raman spectroscopy and TEM studies of CrPS$_4$); Supplementary tables (Lattice parameters of the CuCrP$_2$S$_6$ bulk from DFT calculations and XRD experiments; Atomic concentrations of elements in synthesised CuCrP$_2$S$_6$ bulk crystals obtained in SEM/EDX and TEM/EDX measurements); Supplementary figures (Synthesis scheme used in the present work; Crystallographic structure of CuCrP$_2$S$_6$ in two representations; Representative photos of several CuCrP$_2$S$_6$ bulk crystals used in the present work; Series of XRD patterns measured over $30$ CuCrP$_2$S$_6$ crystals; Crystallographic structure of CuCrP$_2$S$_6$ obtained in the structure determination; EDX elements' distribution maps obtained in TEM measurements of CuCrP$_2$S$_6$ bulk crystal; XPS spectra of CuCrP$_2$S$_6$ (CCPS) bulk crystal collected under laboratory conditions; Results of fit procedure for core-level XPS spectra of CuCrP$_2$S$_6$ bulk crystal collected in the laboratory conditions; resonance spectra measured for CuCrP$_2$S$_6$ in a wide energy range; Schematic representation for processes during resonant photoemission; XRD diffraction spots for CrPS$_4$ bulk crystal; Representative Raman spectrum of bulk CrPS$_4$ crystal; TEM image of the CrPS$_4$ bulk crystal) (PDF).
  
  \item Results of the structure refinement procedure for CuCrP$_2$S$_6$ (CIF).

\end{itemize}

\end{suppinfo}

 
%\bibliography{/Users/YuDedkov/Work/Articles/___REFERENCES___/MatCatEn_Shared_Library.bib}
%\bibliography{/Users/evoloshina/WORK/Articles/_References/MatCatEn_Shared_Library.bib}

%%% REFERENCES

%%% FIGURES

\clearpage

\clearpage
\begin{figure}
\includegraphics[width=\textwidth]{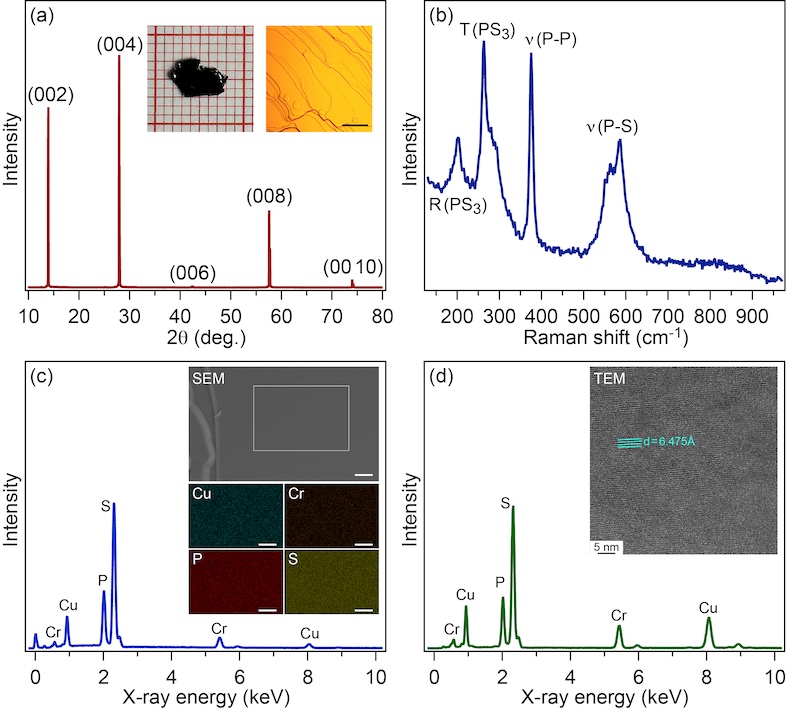}
\caption{\label{fig:CuCrP2S6_XRD_Raman_SEM_TEM} Representative results obtained in charatcteriization of CuCrP$_2$S$_6$ crystals: (a) XRD intensity plot (insets show photo and optical microscopy image of the studied crystals; scale bar is $100\,\mu\mathrm{m}$); (b) Raman spectrum; (c) EDX plot obtained in SEM measurements (insets show the respective SEM and elements' distributions maps; scale bar is $10\,\mu\mathrm{m}$); (d) EDX plot obtained in TEM measurements (insets show the respective high resolution TEM image).}
\end{figure}

\clearpage

\clearpage

\clearpage
\begin{figure}
\includegraphics[width=\textwidth]{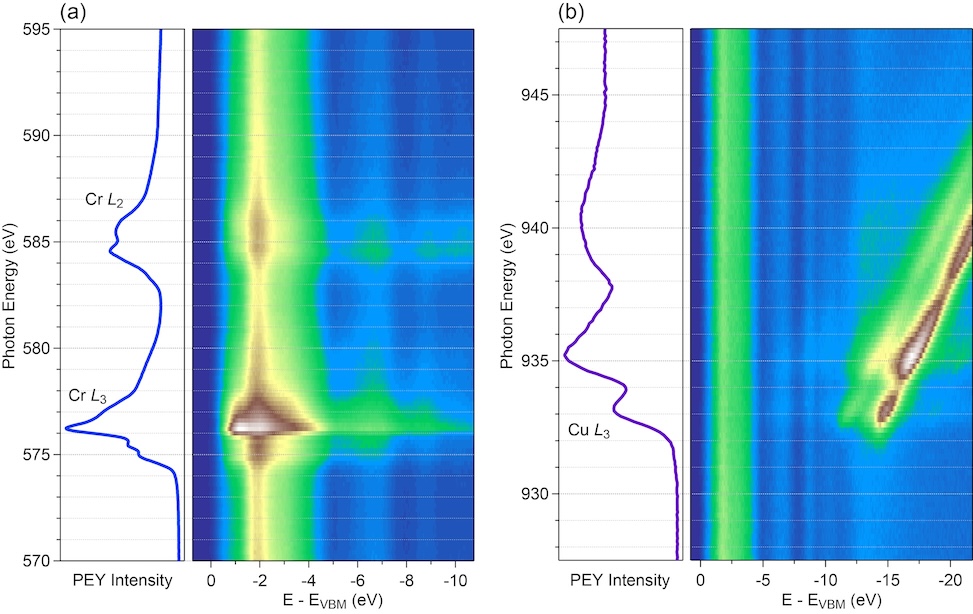}
\caption{\label{fig:CuCrP2S6_ResPES_MAXIV} Valence band photoemission intensity maps $I(E_B, h\nu)$ measured for CuCrP$_2$S$_6$ around the (a) Cr\,$L_{2,3}$ and (b) Cu\,$L_{3}$ absorption edges. The respective NEXAFS reference spectra are presented on the left-hand side of every intensity map.}
\end{figure}

%%%%%%%%%%%%%%%%%%%%%%%%%%%%%%%%%%%%%%%%%%%%%%%%%%%%%%%%%%%%%%%%%%%%%
%% The document title should be given as usual. Some journals require
%% a running title from the author: this should be supplied as an
%% optional argument to \title.
%%%%%%%%%%%%%%%%%%%%%%%%%%%%%%%%%%%%%%%%%%%%%%%%%%%%%%%%%%%%%%%%%%%%%

\clearpage

\noindent
\textbf{Supplementary Material for manuscript:\\ ``Electronic Correlations in Multiferroic van der Waals CuCrP$_2$S$_6$: Insights From X-Ray Spectroscopy and DFT''} %Title of paper

%%%%%%%%%%%%%%%%%%%%%%%%%%%%%%%%%%%%%%%%%%%%%%%%%%%%%%%%%%%%%%%%%%%%%
%% Some journals require a list of abbreviations or keywords to be
%% supplied. These should be set up here, and will be printed after
%% the title and author information, if needed.
%%%%%%%%%%%%%%%%%%%%%%%%%%%%%%%%%%%%%%%%%%%%%%%%%%%%%%%%%%%%%%%%%%%%%
%\abbreviations{IR,NMR,UV}
%\keywords{American Chemical Society, \LaTeX}

%%%%%%%%%%%%%%%%%%%%%%%%%%%%%%%%%%%%%%%%%%%%%%%%%%%%%%%%%%%%%%%%%%%%%
%% The manuscript does not need to include \maketitle, which is
%% executed automatically.
%%%%%%%%%%%%%%%%%%%%%%%%%%%%%%%%%%%%%%%%%%%%%%%%%%%%%%%%%%%%%%%%%%%%%

\section{Supplementary description}

\begin{itemize}

\item Structure refinement for CuCrP$_2$S$_6$

\item Laboratory-based XPS characterization of CuCrP$_2$S$_6$

\item XRD, Raman spectroscopy and TEM studies of CrPS$_4$

\end{itemize}

\section{Supplementary tables}

\begin{itemize}

\item Table\,T1: Lattice parameters of the CuCrP$_2$S$_6$ bulk from DFT calculations and XRD experiments.

\item Table\,T2: Atomic concentrations of elements in synthesised CuCrP$_2$S$_6$ bulk crystals obtained in SEM/EDX and TEM/EDX measurements.

\item Table\,T3: Results obtained for the different magnetic states of 1ML-CuCrP$_2$S$_6$ and bulk CuCrP$_2$S$_6$ with PBE$+U+$D2: $E_\mathrm{tot}$ (in eV per hexagonal unit cell) is the total energy;  magnetic moments of Cr, Cu, P, and S  ($m$, in $\mu_B$).

\end{itemize}

\section{Supplementary figures}

\begin{itemize}

\item Figure\,S1: Synthesis scheme used in the present work: (a) Temperature as a function of time profile used for the CVT synthesis of CuCrP$_2$S$_6$; (b) Schematic drawing of an ampule during CVT growth with arrows indicating the mass ﬂow.

\item Figure\,S2: Crystallographic structure of CuCrP$_2$S$_6$ in two representations used in the present group: (a) $C2/c$ space group and (b) hexagonal unit cell containing $6$ layers.

\item Figure\,S3: DOS plots for bulk CuCrP$_2$S$_6$ AFM ground state obtained with (a) HSE06 functional, (b) PBE$+U$ with $U(\textrm{Cr})=4$\,eV and  $U(\textrm{Cu})=4$\,eV, and (c) PBE$+U$ with $U(\textrm{Cr})=5$\,eV and  $U(\textrm{Cu})=7$\,eV.

\item Figure\,S4: Representative photos of several CuCrP$_2$S$_6$ bulk crystals used in the present work.

\item Figure\,S5: Series of XRD patterns measured over $30$ CuCrP$_2$S$_6$ crystals in the $2\theta$ range corresponding to $(002)$ and $(004)$ diffraction peaks. Vertical dashed lines mark the positions of the respective diffraction peaks for the CrPS$_4$ crystal.

\item Figure\,S6: (a) Crystallographic structure of CuCrP$_2$S$_6$ obtained in the structure determination. Cu1 (Cu1'), Cu2 (Cu2'), and Cu3 (Cu3') denote six Cu positions which occupations were determined at different temperatures. (b) Occupation numbers for Cu1 (Cu1'), Cu2 (Cu2'), and Cu3 (Cu3'). (c) DOS calculated using the crystallographic structure of CuCrP$_2$S$_6$ bulk obtained in the refinement procedure. 

\item Figure\,S7: EDX elements' distribution maps obtained in TEM measurements of CuCrP$_2$S$_6$ bulk crystal.

\item Figure\,S8: XPS spectra of CuCrP$_2$S$_6$ (CCPS) bulk crystal (before and after cleavage) and Cu-poly sample collected under laboratory conditions: (a) survey, (b) Cu\,$2p$, (c) Cr\,$2p$ and Cu\,$LMM$, (d) valence band, (e) P\,$2p$, (f) S\,$2p$. All spectra were collected with photon energy $h\nu=1486.6$\,eV (Al\,$K\alpha$).

\item Figure\,S9: Results of fit procedure for core-level XPS spectra of CuCrP$_2$S$_6$ bulk crystal collected in the laboratory conditions: (a) Cu\,$2p_{3/2}$, (b) P\,$2p$, (c) S\,$2p$.

\item Figure\,S10: ``ON'' and ``OFF'' resonance spectra measured for CuCrP$_2$S$_6$ in a wide energy range at the Cr\,$L_{3}$ absorption edge. Photon energies are marked in the plot.

\item Figure\,S11: Schematic representation for processes during resonant photoemission: (a) direct photoemission channel; (b) core electron excitation process from $2p$ core level on unoccupied $3d^*/4s^*$ states above the Fermi level followed by (c) participator-Auger or (d) spectator-Auger electron decay processes of a resonant excitation.

\item Figure\,S12: XRD diffraction spots for CrPS$_4$ bulk crystal. Inset shows the photo of crystal.

\item Figure\,S13: Representative Raman spectrum of bulk CrPS$_4$ crystal. Peak notations are taken from J. Lee \textit{et al.} Structural and Optical Properties of Single- and Few-Layer Magnetic Semiconductor CrPS$_4$. ACS Nano \textbf{11}, 10935 (2017).

\item Figure\,S14: TEM image of the CrPS$_4$ bulk crystal obtained in the same CVT synthesis. (b) The respective EDX spectrum of CrPS$_4$.

\end{itemize}

\clearpage

\section{Structure refinement for CuCrP$_2$S$_6$}

Our structure refinement procedure for the CuCrP$_2$S$_6$ crystal reveals the $C2/c$ space group in the temperature range from $293$\,K to $\approx150$\,K. For low temperature structure determination at $64$\,K the $Pc$ space group is reported~\cite{Maisonneuve.1993abc}. However, in our DFT analysis the former space group is used giving an adequate description of the studied material at the temperatures used in the present work. At elevated temperatures ($150-293$\,K) the structural motif that is described in the main text could clearly be identified (Fig.\,1(a,b) of the main text), including the disorder of the copper atoms (see Fig.~\ref{fig:CuCrP2S6_StructureRefinement}(a))~\cite{Colombet.1982}. Whilst the occupation of all copper atoms on the disorder positions is approximately equal at room temperature, a clear preference for the ``outer'' positions Cu1 and Cu2 is obtained when lowering the temperature (Fig.~\ref{fig:CuCrP2S6_StructureRefinement}(a,b)). This gradual ordering is in line with the reported structure at $64$\,K in space group type $Pc$~\cite{Maisonneuve.1993abc}. The lattice parameters of CuCrP$_2$S$_6$ crystal obtained in the structure determination are presented in Table~\ref{tab:lattice_parameters} and they are in good agreement with values obteined in the DFT calculations. Moreover, the calculated DOS for CuCrP$_2$S$_6$ using the lattice parameters obtained in the structure refinement experiment gives the similar distribution of the electronic states and value of the band gap ($1.23$\,eV) (Fig.~\ref{fig:CuCrP2S6_StructureRefinement}(c)). We would like to note, that during structure refinement procedure an additional disorder is found in CuCrP$_2$S$_6$ bulk, when P--P dumbbells can be refined to an occupation of $85$\% with additional 15\% total Cu occupation that obtains the same positional disorder as Cu1-3. This is in line with the high residual difference electron density ($4.3\,e^-$) located in between phosphorus ions. The high electron density on the P--P center would correspond to the central disorder positions equivalent to Cu3, which also vanishes in lower-temperature ($<175$\,K) datasets. However, for charge neutrality reasons and due to compositional considerations, refinement of such additional disorder would require another overlay of the modelled sixfold Cu-disorder with $15$\% of a P-P  dumbbell, which is clearly out of scope of the resolution. Therefore, this disorder was omitted, yielding a comparably high residual electron density. However, this type of disorder can be simply considered as a defect and due to the strong covalent and ionic interactions in CuCrP$_2$S$_6$ will not strongly influence the results of the spectroscopic experiments in the present work. All refinement values, as well as additional structural information for all measurements can be obtained from the attached crystallographic information file (CIF).

\clearpage

\section{Laboratory-based XPS characterization of CuCrP$_2$S$_6$}

Initial characterization of the CuCrP$_2$S$_6$ crystals was perfromed under laboratory conditions and results ate are compiled in Fig.~\ref{fig:CuCrP2S6_XPS_SHU_overview}. XPS spectra of the non-cleaved CuCrP$_2$S$_6$ sample (Fig.~\ref{fig:CuCrP2S6_XPS_SHU_overview}(a)) demonstrate the presence of all photoemission lines together with signals which can be assigned to the surface contaminations (O\,$1s$ and C\,$1s$ lines). In laboratory-based XPS measurements, cleaving of the CuCrP$_2$S$_6$ sample results in a clean surface of without O\,$1s$ and C\,$1s$ signals. The positions of the Cu\,$2p$ XPS and the Cu\,$L_3M_{45}M_{45}$ Auger lines confirm the Cu$^{1+}$ charge state~\cite{Wang.2018oja,Liu.2022} (Fig.~\ref{fig:CuCrP2S6_XPS_SHU_overview}(b,c)). On the other hand, the S\,$2p$ and P\,$2p$ XPS lines measured under laboratory conditions do not demonstrate any clear spin-orbit split doublets. A peak-fit analysis of the respective Cu\,$2p_{3/2}$, S\,$2p$ and P\,$2p$ XPS lines presented in  Fig.~\ref{fig:CuCrP2S6_XPS_FIT_Cu2p_P2p_S2p} shows that in all cases the photoemission lines consist of two parts shifted by $\approx0.9$\,eV with respect to each other. This is a result of the large footprint of the Al\,$K\alpha$ X-ray source and the large acceptance area of the analyzer. Indeed, in this case the XPS signal is collected from several areas of the CuCrP$_2$S$_6$ crystal, which might be not in a good electrical contact between them (for example, small single flakes of the crystal formed during cleavage procedure).

\clearpage

\section{XRD, Raman spectroscopy and TEM studies of CrPS$_4$}

The representative XRD plot for CrPS$_4$ crystal is shown in Fig.~\ref{fig:CrPS4_XRD}. According to the $C2/m$ group symmetry of CrPS$_4$, the respective $2\theta$ values for $(001)$ and $(002)$ diffraction peaks are $14.789^\circ$ and $29.433^\circ$, which are quite distinct from the corresponding values for CuCrP$_2$S$_6$ crystals (see Figs.~\ref{fig:CuCrP2S6_XRD_30samples} and \ref{fig:CrPS4_XRD}). Also the relative intensities of these peaks for CrPS$_4$ correspond to the $C2/m$ group~\cite{Hou.2020,Jin.2022yb} and different from those for CuCrP$_2$S$_6$, which has $C2/c$ symmetry group.

The Raman spectrum of CrPS$_4$ bulk crystal ($C2/m$ space group) comprises of peaks' array presented in Fig.~\ref{fig:CrPS4_Raman} and these peaks can be attributed to a combination of $10$ \textit{A} and $8$ \textit{B} Brillouin zone centered modes, belonging to the $C2$ point group~\cite{Kim.2021asd}. The presented Raman spectra for CuCrP$_2$S$_6$ and CrPS$_4$ are characteristic for these crystals~\cite{Susner.2020,Io.2023,Rao.2023,Ma.2023,Lee.2017f9o,Kim.2021asd,Wu.2019wly} and can be used as a clear fingerprint to discriminate between the two compounds, which are usually obtained during CVT synthesis.

The comparison of TEM images of CuCrP$_2$S$_6$ and CrPS$_4$ (see Fig.~\ref{fig:CrPS4_TEM_EDX}) demonstrates a more ordered structure in the later system. It is found that CuCrP$_2$S$_6$ crystals (compared to CrPS$_4$) are very sensitive to the high-energy electron beam used in TEM experiments, that can damage the crystal lattice and locally change the composition. It could be explained by the more ionic character of the chemical bonds in CrPS$_4$, rendering these crystals more stable under electron beam irradiation. The distances between single planes in CrPS$_4$ are $6.105$\,\AA, as extracted from TEM images, while the EDX analysis gives a very good stoichiometry for these crystals. (The small concentration of Cu detected in both SEM/EDX and TEM/EDX measurements on CrPS$_4$ is due to the fact that both compounds are synthesised in the same process.)

\clearpage
%\bibliography{/Users/YuDedkov/Work/Articles/___REFERENCES___/MatCatEn_Shared_Library.bib}

\begin{mcitethebibliography}{55}
\providecommand*\natexlab[1]{#1}
\providecommand*\mciteSetBstSublistMode[1]{}
\providecommand*\mciteSetBstMaxWidthForm[2]{}
\providecommand*\mciteBstWouldAddEndPuncttrue
  {\def\EndOfBibitem{\unskip.}}
\providecommand*\mciteBstWouldAddEndPunctfalse
  {\let\EndOfBibitem\relax}
\providecommand*\mciteSetBstMidEndSepPunct[3]{}
\providecommand*\mciteSetBstSublistLabelBeginEnd[3]{}
\providecommand*\EndOfBibitem{}
\mciteSetBstSublistMode{f}
\mciteSetBstMaxWidthForm{subitem}{(\alph{mcitesubitemcount})}
\mciteSetBstSublistLabelBeginEnd
  {\mcitemaxwidthsubitemform\space}
  {\relax}
  {\relax}

\bibitem[Novoselov \latin{et~al.}(2005)Novoselov, Geim, Morozov, Jiang,
  Katsnelson, Grigorieva, Dubonos, and Firsov]{Novoselov.2005}
Novoselov,~K.~S.; Geim,~A.~K.; Morozov,~S.~V.; Jiang,~D.; Katsnelson,~M.~I.;
  Grigorieva,~I.~V.; Dubonos,~S.~V.; Firsov,~A.~A. {Two-dimensional gas of
  massless Dirac fermions in graphene}. \emph{Nature} \textbf{2005},
  \emph{438}, 197--200\relax
\mciteBstWouldAddEndPuncttrue
\mciteSetBstMidEndSepPunct{\mcitedefaultmidpunct}
{\mcitedefaultendpunct}{\mcitedefaultseppunct}\relax
\EndOfBibitem
\bibitem[Zhang \latin{et~al.}(2005)Zhang, Tan, Stormer, and Kim]{Zhang.2005}
Zhang,~Y.; Tan,~Y.-W.; Stormer,~H.~L.; Kim,~P. {Experimental observation of the
  quantum Hall effect and Berry's phase in graphene}. \emph{Nature}
  \textbf{2005}, \emph{438}, 201--204\relax
\mciteBstWouldAddEndPuncttrue
\mciteSetBstMidEndSepPunct{\mcitedefaultmidpunct}
{\mcitedefaultendpunct}{\mcitedefaultseppunct}\relax
\EndOfBibitem
\bibitem[Manzeli \latin{et~al.}(2017)Manzeli, Ovchinnikov, Pasquier, Yazyev,
  and Kis]{Manzeli.2017}
Manzeli,~S.; Ovchinnikov,~D.; Pasquier,~D.; Yazyev,~O.~V.; Kis,~A. {2D
  transition metal dichalcogenides}. \emph{Nat. Rev. Mater.} \textbf{2017},
  \emph{2}, 17033\relax
\mciteBstWouldAddEndPuncttrue
\mciteSetBstMidEndSepPunct{\mcitedefaultmidpunct}
{\mcitedefaultendpunct}{\mcitedefaultseppunct}\relax
\EndOfBibitem
\bibitem[Geim and Grigorieva(2013)Geim, and Grigorieva]{Geim.2013}
Geim,~A.~K.; Grigorieva,~I.~V. {Van der Waals heterostructures}. \emph{Nature}
  \textbf{2013}, \emph{499}, 419--425\relax
\mciteBstWouldAddEndPuncttrue
\mciteSetBstMidEndSepPunct{\mcitedefaultmidpunct}
{\mcitedefaultendpunct}{\mcitedefaultseppunct}\relax
\EndOfBibitem
\bibitem[Yu \latin{et~al.}(2014)Yu, Gorbachev, Tu, Kretinin, Cao, Jalil,
  Withers, Ponomarenko, Piot, Potemski, Elias, Chen, Watanabe, Taniguchi,
  Grigorieva, Novoselov, Fal’ko, Geim, and Mishchenko]{Yu.2014vu}
Yu,~G.~L.; Gorbachev,~R.~V.; Tu,~J.~S.; Kretinin,~A.~V.; Cao,~Y.; Jalil,~R.;
  Withers,~F.; Ponomarenko,~L.~A.; Piot,~B.~A.; Potemski,~M. \latin{et~al.}
  {Hierarchy of Hofstadter states and replica quantum Hall ferromagnetism in
  graphene superlattices}. \emph{Nat. Phys.} \textbf{2014}, \emph{10},
  525--529\relax
\mciteBstWouldAddEndPuncttrue
\mciteSetBstMidEndSepPunct{\mcitedefaultmidpunct}
{\mcitedefaultendpunct}{\mcitedefaultseppunct}\relax
\EndOfBibitem
\bibitem[Liu \latin{et~al.}(2016)Liu, Weiss, Duan, Cheng, Huang, and
  Duan]{Liu.2016xab}
Liu,~Y.; Weiss,~N.~O.; Duan,~X.; Cheng,~H.-C.; Huang,~Y.; Duan,~X. {Van der
  Waals heterostructures and devices}. \emph{Nat. Rev. Mater.} \textbf{2016},
  \emph{1}, 16042\relax
\mciteBstWouldAddEndPuncttrue
\mciteSetBstMidEndSepPunct{\mcitedefaultmidpunct}
{\mcitedefaultendpunct}{\mcitedefaultseppunct}\relax
\EndOfBibitem
\bibitem[Yankowitz \latin{et~al.}(2019)Yankowitz, Ma, Jarillo-Herrero, and
  LeRoy]{Yankowitz.2019}
Yankowitz,~M.; Ma,~Q.; Jarillo-Herrero,~P.; LeRoy,~B.~J. {van der Waals
  heterostructures combining graphene and hexagonal boron nitride}. \emph{Nat.
  Rev. Phys.} \textbf{2019}, \emph{1}, 112--125\relax
\mciteBstWouldAddEndPuncttrue
\mciteSetBstMidEndSepPunct{\mcitedefaultmidpunct}
{\mcitedefaultendpunct}{\mcitedefaultseppunct}\relax
\EndOfBibitem
\bibitem[Pham \latin{et~al.}(2022)Pham, Bodepudi, Shehzad, Liu, Xu, Yu, and
  Duan]{Pham.2022}
Pham,~P.~V.; Bodepudi,~S.~C.; Shehzad,~K.; Liu,~Y.; Xu,~Y.; Yu,~B.; Duan,~X.
  {2D heterostructures for ubiquitous electronics and optoelectronics:
  Principles, opportunities, and challenges}. \emph{Chem. Rev.} \textbf{2022},
  \emph{122}, 6514--6613\relax
\mciteBstWouldAddEndPuncttrue
\mciteSetBstMidEndSepPunct{\mcitedefaultmidpunct}
{\mcitedefaultendpunct}{\mcitedefaultseppunct}\relax
\EndOfBibitem
\bibitem[Samal \latin{et~al.}(2020)Samal, Sanyal, Chakraborty, and
  Rout]{Samal.2020}
Samal,~R.; Sanyal,~G.; Chakraborty,~B.; Rout,~C.~S. {Two-dimensional transition
  metal phosphorous trichalcogenides (MPX$_3$): A review on emerging trends,
  current state and future perspectives}. \emph{J. Mater. Chem. A}
  \textbf{2020}, \emph{9}, 2560--2591\relax
\mciteBstWouldAddEndPuncttrue
\mciteSetBstMidEndSepPunct{\mcitedefaultmidpunct}
{\mcitedefaultendpunct}{\mcitedefaultseppunct}\relax
\EndOfBibitem
\bibitem[Zhang \latin{et~al.}(2023)Zhang, Wei, Qiu, Zhang, Liu, Hu, Luo, and
  Liu]{Zhang.2023xyz}
Zhang,~H.; Wei,~T.; Qiu,~Y.; Zhang,~S.; Liu,~Q.; Hu,~G.; Luo,~J.; Liu,~X.
  {Recent progress in metal phosphorous chalcogenides: Potential
  high‐performance electrocatalysts}. \emph{Small} \textbf{2023}, \emph{19},
  e2207249\relax
\mciteBstWouldAddEndPuncttrue
\mciteSetBstMidEndSepPunct{\mcitedefaultmidpunct}
{\mcitedefaultendpunct}{\mcitedefaultseppunct}\relax
\EndOfBibitem
\bibitem[Du \latin{et~al.}(2016)Du, Wang, Liu, Hu, Utama, Gan, Xiong, and
  Kloc]{Du.2016}
Du,~K.-z.; Wang,~X.-z.; Liu,~Y.; Hu,~P.; Utama,~M. I.~B.; Gan,~C.~K.;
  Xiong,~Q.; Kloc,~C. {Weak van der Waals stacking, wide-range band gap, and
  Raman study on ultrathin layers of metal phosphorus trichalcogenides}.
  \emph{ACS Nano} \textbf{2016}, \emph{10}, 1738--1743\relax
\mciteBstWouldAddEndPuncttrue
\mciteSetBstMidEndSepPunct{\mcitedefaultmidpunct}
{\mcitedefaultendpunct}{\mcitedefaultseppunct}\relax
\EndOfBibitem
\bibitem[Dedkov \latin{et~al.}(2023)Dedkov, Guo, and Voloshina]{Dedkov.2023ues}
Dedkov,~Y.; Guo,~Y.; Voloshina,~E. {Progress in the studies of electronic and
  magnetic properties of layered MPX$_3$ materials (M: transition metal, X:
  chalcogen)}. \emph{Electron. Struct.} \textbf{2023}, \emph{5}, 043001\relax
\mciteBstWouldAddEndPuncttrue
\mciteSetBstMidEndSepPunct{\mcitedefaultmidpunct}
{\mcitedefaultendpunct}{\mcitedefaultseppunct}\relax
\EndOfBibitem
\bibitem[Kamata \latin{et~al.}(1997)Kamata, Noguchi, Suzuki, Tezuka,
  Kashiwakura, Ohno, and Nakai]{Kamata.1997}
Kamata,~A.; Noguchi,~K.; Suzuki,~K.; Tezuka,~H.; Kashiwakura,~T.; Ohno,~Y.;
  Nakai,~S.-i. {Resonant $2p\rightarrow 3d$ photoemission measurement of MPS$_3$ (M = Mn,
  Fe, Ni)}. \emph{J. Phys. Soc. Jpn.} \textbf{1997}, \emph{66}, 401--407\relax
\mciteBstWouldAddEndPuncttrue
\mciteSetBstMidEndSepPunct{\mcitedefaultmidpunct}
{\mcitedefaultendpunct}{\mcitedefaultseppunct}\relax
\EndOfBibitem
\bibitem[Yang \latin{et~al.}(2020)Yang, Zhou, Guo, Dedkov, and
  Voloshina]{Yang.2020uc}
Yang,~J.; Zhou,~Y.; Guo,~Q.; Dedkov,~Y.; Voloshina,~E. {Electronic, magnetic
  and optical properties of MnPX$_3$ (X = S, Se) monolayers with and without
  chalcogen defects: a first-principles study}. \emph{RSC Adv.} \textbf{2020},
  \emph{10}, 851--864\relax
\mciteBstWouldAddEndPuncttrue
\mciteSetBstMidEndSepPunct{\mcitedefaultmidpunct}
{\mcitedefaultendpunct}{\mcitedefaultseppunct}\relax
\EndOfBibitem
\bibitem[Jin \latin{et~al.}(2022)Jin, Yan, Kremer, Voloshina, and
  Dedkov]{Jin.2022}
Jin,~Y.; Yan,~M.; Kremer,~T.; Voloshina,~E.; Dedkov,~Y. {Mott–Hubbard
  insulating state for the layered van der Waals FePX$_3$ (X: S, Se) as revealed
  by NEXAFS and resonant photoelectron spectroscopy}. \emph{Sci. Rep.}
  \textbf{2022}, \emph{12}, 735\relax
\mciteBstWouldAddEndPuncttrue
\mciteSetBstMidEndSepPunct{\mcitedefaultmidpunct}
{\mcitedefaultendpunct}{\mcitedefaultseppunct}\relax
\EndOfBibitem
\bibitem[Kim \latin{et~al.}(2018)Kim, Kim, Sandilands, Sinn, Lee, Son, Lee,
  Choi, Kim, Park, Jeon, Kim, Park, Park, Moon, and Noh]{Kim.2018}
Kim,~S.~Y.; Kim,~T.~Y.; Sandilands,~L.~J.; Sinn,~S.; Lee,~M.-C.; Son,~J.;
  Lee,~S.; Choi,~K.-Y.; Kim,~W.; Park,~B.-G. \latin{et~al.}  {Charge-spin
  correlation in van der Waals antiferromagnet NiPS$_3$}. \emph{Phys. Rev.
  Lett.} \textbf{2018}, \emph{120}, 136402\relax
\mciteBstWouldAddEndPuncttrue
\mciteSetBstMidEndSepPunct{\mcitedefaultmidpunct}
{\mcitedefaultendpunct}{\mcitedefaultseppunct}\relax
\EndOfBibitem
\bibitem[Yan \latin{et~al.}(2021)Yan, Jin, Wu, Tsaturyan, Makarova, Smirnov,
  Voloshina, and Dedkov]{Yan.2021ni}
Yan,~M.; Jin,~Y.; Wu,~Z.; Tsaturyan,~A.; Makarova,~A.; Smirnov,~D.;
  Voloshina,~E.; Dedkov,~Y. {Correlations in the electronic structure of van
  der Waals NiPS$_3$ crystals: An x‑ray absorption and resonant photoelectron
  spectroscopy study}. \emph{J. Phys. Chem. Lett.} \textbf{2021}, \emph{12},
  2400--2405\relax
\mciteBstWouldAddEndPuncttrue
\mciteSetBstMidEndSepPunct{\mcitedefaultmidpunct}
{\mcitedefaultendpunct}{\mcitedefaultseppunct}\relax
\EndOfBibitem
\bibitem[Jin \latin{et~al.}(2022)Jin, Jin, Li, Yan, Guo, Zhou, Preobrajenski,
  Dedkov, and Voloshina]{Jin.2022jc}
Jin,~Y.; Jin,~Y.; Li,~K.; Yan,~M.; Guo,~Y.; Zhou,~Y.; Preobrajenski,~A.;
  Dedkov,~Y.; Voloshina,~E. {Mixed insulating state for van der Waals CoPS$_3$}.
  \emph{J. Phys. Chem. Lett.} \textbf{2022}, \emph{13}, 10486--10493\relax
\mciteBstWouldAddEndPuncttrue
\mciteSetBstMidEndSepPunct{\mcitedefaultmidpunct}
{\mcitedefaultendpunct}{\mcitedefaultseppunct}\relax
\EndOfBibitem
\bibitem[Li \latin{et~al.}(2023)Li, Yan, Jin, Jin, Guo, Voloshina, and
  Dedkov]{Li.2023}
Li,~K.; Yan,~M.; Jin,~Y.; Jin,~Y.; Guo,~Y.; Voloshina,~E.; Dedkov,~Y. {Dual
  character of the insulating state in the van der Waals Fe$_{1-x}$Ni$_x$PS$_3$ alloyed
  compounds.} \emph{J. Phys. Chem. Lett.} \textbf{2023}, \emph{14},
  57--65\relax
\mciteBstWouldAddEndPuncttrue
\mciteSetBstMidEndSepPunct{\mcitedefaultmidpunct}
{\mcitedefaultendpunct}{\mcitedefaultseppunct}\relax
\EndOfBibitem
\bibitem[Yan \latin{et~al.}(2023)Yan, Jin, Voloshina, and Dedkov]{Yan.202396s}
Yan,~M.; Jin,~Y.; Voloshina,~E.; Dedkov,~Y. {Electronic Correlations in Fe$_x$Ni$_y$PS$_3$ van der Waals materials: Insights from angle-resolved photoelectron
  spectroscopy and DFT}. \emph{J. Phys. Chem. Lett.} \textbf{2023}, \emph{14},
  9774--9779\relax
\mciteBstWouldAddEndPuncttrue
\mciteSetBstMidEndSepPunct{\mcitedefaultmidpunct}
{\mcitedefaultendpunct}{\mcitedefaultseppunct}\relax
\EndOfBibitem
\bibitem[Voloshina \latin{et~al.}(2023)Voloshina, Jin, and
  Dedkov]{Voloshina.2023}
Voloshina,~E.; Jin,~Y.; Dedkov,~Y. {ARPES studies of the ground state
  electronic properties of the van der Waals transition metal trichalcogenide
  CoPS$_3$}. \emph{Chem. Phys. Lett.} \textbf{2023}, \emph{823}, 140511\relax
\mciteBstWouldAddEndPuncttrue
\mciteSetBstMidEndSepPunct{\mcitedefaultmidpunct}
{\mcitedefaultendpunct}{\mcitedefaultseppunct}\relax
\EndOfBibitem
\bibitem[Koitzsch \latin{et~al.}(2023)Koitzsch, Klaproth, Selter, Shemerliuk,
  Aswartham, Janson, Büchner, and Knupfer]{Koitzsch.2023}
Koitzsch,~A.; Klaproth,~T.; Selter,~S.; Shemerliuk,~Y.; Aswartham,~S.;
  Janson,~O.; Büchner,~B.; Knupfer,~M. {Intertwined electronic and magnetic
  structure of the van-der-Waals antiferromagnet Fe$_2$P$_2$S$_6$}. \emph{npj Quantum
  Mater.} \textbf{2023}, \emph{8}, 27\relax
\mciteBstWouldAddEndPuncttrue
\mciteSetBstMidEndSepPunct{\mcitedefaultmidpunct}
{\mcitedefaultendpunct}{\mcitedefaultseppunct}\relax
\EndOfBibitem
\bibitem[Nitschke \latin{et~al.}(2023)Nitschke, Esteras, Gutnikov, Schiller,
  Manas, Coronado, Stupar, Zamborlini, Ponzoni, Baldoví, and
  Cinchetti]{Nitschke.2023}
Nitschke,~J.~E.; Esteras,~D.~L.; Gutnikov,~M.; Schiller,~K.; Manas,~S.;
  Coronado,~E.; Stupar,~M.; Zamborlini,~G.; Ponzoni,~S.; Baldoví,~J.~J.
  \latin{et~al.}  {Valence band electronic structure of the van der Waals
  antiferromagnet FePS$_3$}. \emph{Mater. Today Electron.} \textbf{2023},
  100061\relax
\mciteBstWouldAddEndPuncttrue
\mciteSetBstMidEndSepPunct{\mcitedefaultmidpunct}
{\mcitedefaultendpunct}{\mcitedefaultseppunct}\relax
\EndOfBibitem
\bibitem[Strasdas \latin{et~al.}(2023)Strasdas, Pestka, Rybak, Budniak, Leuth,
  Boban, Feyer, Cojocariu, Baranowski, Avila, Dudin, Bostwick, Jozwiak,
  Rotenberg, Autieri, Amouyal, Plucinski, Lifshitz, Birowska, and
  Morgenstern]{Strasdas.2023}
Strasdas,~J.; Pestka,~B.; Rybak,~M.; Budniak,~A.~K.; Leuth,~N.; Boban,~H.;
  Feyer,~V.; Cojocariu,~I.; Baranowski,~D.; Avila,~J. \latin{et~al.}
  {Electronic band structure changes across the antiferromagnetic phase
  transition of exfoliated MnPS$_3$ flakes probed by $\mu$-ARPES}. \emph{Nano Lett.}
  \textbf{2023}, \emph{23}, 10342--10349\relax
\mciteBstWouldAddEndPuncttrue
\mciteSetBstMidEndSepPunct{\mcitedefaultmidpunct}
{\mcitedefaultendpunct}{\mcitedefaultseppunct}\relax
\EndOfBibitem
\bibitem[Klaproth \latin{et~al.}(2023)Klaproth, Aswartham, Shemerliuk, Selter,
  Janson, Brink, Büchner, Knupfer, Pazek, Mikhailova, Efimenko, Hayn,
  Savoyant, Gubanov, and Koitzsch]{Klaproth.2023}
Klaproth,~T.; Aswartham,~S.; Shemerliuk,~Y.; Selter,~S.; Janson,~O.; Brink,~J.
  v.~d.; Büchner,~B.; Knupfer,~M.; Pazek,~S.; Mikhailova,~D. \latin{et~al.}
  {Origin of the magnetic exciton in the van der Waals antiferromagnet NiPS$_3$}.
  \emph{Phys. Rev. Lett.} \textbf{2023}, \emph{131}, 256504\relax
\mciteBstWouldAddEndPuncttrue
\mciteSetBstMidEndSepPunct{\mcitedefaultmidpunct}
{\mcitedefaultendpunct}{\mcitedefaultseppunct}\relax
\EndOfBibitem
\bibitem[Du \latin{et~al.}(2018)Du, Liang, Dangol, Zhao, Ren, Madhavi, and
  Yan]{Du.2018}
Du,~C.-F.; Liang,~Q.; Dangol,~R.; Zhao,~J.; Ren,~H.; Madhavi,~S.; Yan,~Q.
  {Layered trichalcogenidophosphate: A new catalyst family for water
  splitting}. \emph{Nano-Micro. Lett.} \textbf{2018}, \emph{10}, 67\relax
\mciteBstWouldAddEndPuncttrue
\mciteSetBstMidEndSepPunct{\mcitedefaultmidpunct}
{\mcitedefaultendpunct}{\mcitedefaultseppunct}\relax
\EndOfBibitem
\bibitem[Wang \latin{et~al.}(2018)Wang, Shifa, Yu, He, Liu, Wang, Wang, Zhan,
  Lou, Xia, and He]{Wang.2018flk}
Wang,~F.; Shifa,~T.~A.; Yu,~P.; He,~P.; Liu,~Y.; Wang,~F.; Wang,~Z.; Zhan,~X.;
  Lou,~X.; Xia,~F. \latin{et~al.}  {New frontiers on van der Waals layered
  metal phosphorous trichalcogenides}. \emph{Adv. Funct. Mater.} \textbf{2018},
  \emph{28}, 1802151\relax
\mciteBstWouldAddEndPuncttrue
\mciteSetBstMidEndSepPunct{\mcitedefaultmidpunct}
{\mcitedefaultendpunct}{\mcitedefaultseppunct}\relax
\EndOfBibitem
\bibitem[Susner \latin{et~al.}(2020)Susner, Rao, Pelton, McLeod, and
  Maruyama]{Susner.2020}
Susner,~M.~A.; Rao,~R.; Pelton,~A.~T.; McLeod,~M.~V.; Maruyama,~B.
  {Temperature-dependent Raman scattering and x-ray diffraction study of phase
  transitions in layered multiferroic CuCrP$_2$S$_6$}. \emph{Phys. Rev. Mater.}
  \textbf{2020}, \emph{4}, 104003\relax
\mciteBstWouldAddEndPuncttrue
\mciteSetBstMidEndSepPunct{\mcitedefaultmidpunct}
{\mcitedefaultendpunct}{\mcitedefaultseppunct}\relax
\EndOfBibitem
\bibitem[Selter \latin{et~al.}(2023)Selter, Bestha, Bhattacharyya, Özer,
  Shemerliuk, Roslova, Vinokurova, Corredor, Veyrat, Wolter, Hozoi, Büchner,
  and Aswartham]{Selter.2023}
Selter,~S.; Bestha,~K.~K.; Bhattacharyya,~P.; Özer,~B.; Shemerliuk,~Y.;
  Roslova,~M.; Vinokurova,~E.; Corredor,~L.~T.; Veyrat,~L.; Wolter,~A. U.~B.
  \latin{et~al.}  {Crystal growth, exfoliation, and magnetic properties of
  quaternary quasi-two-dimensional CuCrP$_2$S$_6$}. \emph{Phys. Rev. Mater.}
  \textbf{2023}, \emph{7}, 033402\relax
\mciteBstWouldAddEndPuncttrue
\mciteSetBstMidEndSepPunct{\mcitedefaultmidpunct}
{\mcitedefaultendpunct}{\mcitedefaultseppunct}\relax
\EndOfBibitem
\bibitem[Wang \latin{et~al.}(2023)Wang, Shang, Zhang, Kang, Liu, Wang, Chen,
  Liu, Tang, Zeng, Guo, Cheng, Liu, Pan, Tong, Wu, Xie, Wang, Deng, Zhai, Deng,
  Hong, and Zhao]{Wang.2023kp}
Wang,~X.; Shang,~Z.; Zhang,~C.; Kang,~J.; Liu,~T.; Wang,~X.; Chen,~S.; Liu,~H.;
  Tang,~W.; Zeng,~Y.-J. \latin{et~al.}  {Electrical and magnetic anisotropies
  in van der Waals multiferroic CuCrP$_2$S$_6$}. \emph{Nat. Commun.} \textbf{2023},
  \emph{14}, 840\relax
\mciteBstWouldAddEndPuncttrue
\mciteSetBstMidEndSepPunct{\mcitedefaultmidpunct}
{\mcitedefaultendpunct}{\mcitedefaultseppunct}\relax
\EndOfBibitem
\bibitem[Park \latin{et~al.}(2022)Park, Shahee, Kim, Patil, Guda,
  Ter‐Oganessian, and Kim]{Park.2022}
Park,~C.~B.; Shahee,~A.; Kim,~K.; Patil,~D.~R.; Guda,~S.~A.;
  Ter‐Oganessian,~N.; Kim,~K.~H. {Observation of spin‐induced
  ferroelectricity in a layered van der Waals antiferromagnet CuCrP$_2$S$_6$}.
  \emph{Adv. Electron. Mater.} \textbf{2022}, \emph{8}, 2101072\relax
\mciteBstWouldAddEndPuncttrue
\mciteSetBstMidEndSepPunct{\mcitedefaultmidpunct}
{\mcitedefaultendpunct}{\mcitedefaultseppunct}\relax
\EndOfBibitem
\bibitem[Pei \latin{et~al.}(2016)Pei, Luo, Lin, Song, Hu, Zou, Yu, Tong, Song,
  Lu, and Sun]{Pei.2016}
Pei,~Q.~L.; Luo,~X.; Lin,~G.~T.; Song,~J.~Y.; Hu,~L.; Zou,~Y.~M.; Yu,~L.;
  Tong,~W.; Song,~W.~H.; Lu,~W.~J.; Sun,~Y.~P. {Spin dynamics, electronic, and
  thermal transport properties of two-dimensional CrPS$_4$ single crystal}.
  \emph{J. Appl. Phys.} \textbf{2016}, \emph{119}, 043902\relax
\mciteBstWouldAddEndPuncttrue
\mciteSetBstMidEndSepPunct{\mcitedefaultmidpunct}
{\mcitedefaultendpunct}{\mcitedefaultseppunct}\relax
\EndOfBibitem
\bibitem[Lee \latin{et~al.}(2017)Lee, Ko, Kim, Bark, Kang, Jung, Park, Lee,
  Ryu, and Lee]{Lee.2017f9o}
Lee,~J.; Ko,~T.~Y.; Kim,~J.~H.; Bark,~H.; Kang,~B.; Jung,~S.-G.; Park,~T.;
  Lee,~Z.; Ryu,~S.; Lee,~C. {Structural and optical properties of single- and
  few-layer magnetic semiconductor CrPS$_4$}. \emph{ACS Nano} \textbf{2017},
  \emph{11}, 10935--10944\relax
\mciteBstWouldAddEndPuncttrue
\mciteSetBstMidEndSepPunct{\mcitedefaultmidpunct}
{\mcitedefaultendpunct}{\mcitedefaultseppunct}\relax
\EndOfBibitem
\bibitem[Susilo \latin{et~al.}(2020)Susilo, Jang, Feng, Du, Yan, Dong, Yuan,
  Petrovic, Shim, Kim, and Chen]{Susilo.2020}
Susilo,~R.~A.; Jang,~B.~G.; Feng,~J.; Du,~Q.; Yan,~Z.; Dong,~H.; Yuan,~M.;
  Petrovic,~C.; Shim,~J.~H.; Kim,~D.~Y.; Chen,~B. {Band gap crossover and
  insulator-metal transition in the compressed layered CrPS$_4$}.
  \emph{npj Quantum Mater.} \textbf{2020}, \emph{5}, 58\relax
\mciteBstWouldAddEndPuncttrue
\mciteSetBstMidEndSepPunct{\mcitedefaultmidpunct}
{\mcitedefaultendpunct}{\mcitedefaultseppunct}\relax
\EndOfBibitem
\bibitem[Dolomanov \latin{et~al.}(2009)Dolomanov, Bourhis, Gildea, Howard, and
  Puschmann]{Dolomanov.2009}
Dolomanov,~O.~V.; Bourhis,~L.~J.; Gildea,~R.~J.; Howard,~J. A.~K.;
  Puschmann,~H. {OLEX2: a complete structure solution, refinement and analysis
  program}. \emph{J. Appl. Crystallogr.} \textbf{2009}, \emph{42},
  339--341\relax
\mciteBstWouldAddEndPuncttrue
\mciteSetBstMidEndSepPunct{\mcitedefaultmidpunct}
{\mcitedefaultendpunct}{\mcitedefaultseppunct}\relax
\EndOfBibitem
\bibitem[Sheldrick(2015)]{Sheldrick.2015}
Sheldrick,~G.~M. {SHELXT - Integrated space‐group and crystal‐structure
  determination}. \emph{Acta Crystallogr. Sect. A} \textbf{2015}, \emph{71},
  3--8\relax
\mciteBstWouldAddEndPuncttrue
\mciteSetBstMidEndSepPunct{\mcitedefaultmidpunct}
{\mcitedefaultendpunct}{\mcitedefaultseppunct}\relax
\EndOfBibitem
\bibitem[Sheldrick(2015)]{Sheldrick.2015kit}
Sheldrick,~G.~M. {Crystal structure refinement with SHELXL}. \emph{Acta
  Crystallogr. Sect. C} \textbf{2015}, \emph{71}, 3--8\relax
\mciteBstWouldAddEndPuncttrue
\mciteSetBstMidEndSepPunct{\mcitedefaultmidpunct}
{\mcitedefaultendpunct}{\mcitedefaultseppunct}\relax
\EndOfBibitem
\bibitem[Preobrajenski \latin{et~al.}(2023)Preobrajenski, Generalov, Öhrwall,
  Tchaplyguine, Tarawneh, Appelfeller, Frampton, and Walsh]{Preobrajenski.2023}
Preobrajenski,~A.; Generalov,~A.; Öhrwall,~G.; Tchaplyguine,~M.; Tarawneh,~H.;
  Appelfeller,~S.; Frampton,~E.; Walsh,~N. {FlexPES: a versatile soft X‐ray
  beamline at MAX IV Laboratory}. \emph{J. Synchrotron Radiat.} \textbf{2023},
  \emph{30}, 831--840\relax
\mciteBstWouldAddEndPuncttrue
\mciteSetBstMidEndSepPunct{\mcitedefaultmidpunct}
{\mcitedefaultendpunct}{\mcitedefaultseppunct}\relax
\EndOfBibitem
\bibitem[Kresse and Furthmüller(1996)Kresse, and Furthmüller]{Kresse.1996uh}
Kresse,~G.; Furthmüller,~J. {Efficient iterative schemes for ab initio
  total-energy calculations using a plane-wave basis set}. \emph{Phys. Rev. B}
  \textbf{1996}, \emph{54}, 11169--11186\relax
\mciteBstWouldAddEndPuncttrue
\mciteSetBstMidEndSepPunct{\mcitedefaultmidpunct}
{\mcitedefaultendpunct}{\mcitedefaultseppunct}\relax
\EndOfBibitem
\bibitem[Kresse and Joubert(1999)Kresse, and Joubert]{Kresse.1999}
Kresse,~G.; Joubert,~D. {From ultrasoft pseudopotentials to the projector
  augmented-wave method}. \emph{Phys. Rev. B} \textbf{1999}, \emph{59},
  1758--1775\relax
\mciteBstWouldAddEndPuncttrue
\mciteSetBstMidEndSepPunct{\mcitedefaultmidpunct}
{\mcitedefaultendpunct}{\mcitedefaultseppunct}\relax
\EndOfBibitem
\bibitem[Perdew \latin{et~al.}(1996)Perdew, Burke, and Ernzerhof]{Perdew.1996}
Perdew,~J.~P.; Burke,~K.; Ernzerhof,~M. {Generalized gradient approximation
  made simple}. \emph{Phys. Rev. Lett.} \textbf{1996}, \emph{77},
  3865--3868\relax
\mciteBstWouldAddEndPuncttrue
\mciteSetBstMidEndSepPunct{\mcitedefaultmidpunct}
{\mcitedefaultendpunct}{\mcitedefaultseppunct}\relax
\EndOfBibitem
\bibitem[Blöchl(1994)]{Blochl.1994}
Blöchl,~P.~E. {Projector augmented-wave method}. \emph{Phys. Rev. B}
  \textbf{1994}, \emph{50}, 17953--17979\relax
\mciteBstWouldAddEndPuncttrue
\mciteSetBstMidEndSepPunct{\mcitedefaultmidpunct}
{\mcitedefaultendpunct}{\mcitedefaultseppunct}\relax
\EndOfBibitem
\bibitem[Blöchl \latin{et~al.}(1994)Blöchl, Jepsen, and
  Andersen]{Blochl.1994abc}
Blöchl,~P.~E.; Jepsen,~O.; Andersen,~O.~K. {Improved tetrahedron method for
  Brillouin-zone integrations}. \emph{Physical Review B} \textbf{1994},
  \emph{49}, 16223--16233\relax
\mciteBstWouldAddEndPuncttrue
\mciteSetBstMidEndSepPunct{\mcitedefaultmidpunct}
{\mcitedefaultendpunct}{\mcitedefaultseppunct}\relax
\EndOfBibitem
\bibitem[Grimme(2006)]{Grimme.2006}
Grimme,~S. {Semiempirical GGA‐type density functional constructed with a
  long‐range dispersion correction}. \emph{J. Comput. Chem.} \textbf{2006},
  \emph{27}, 1787--1799\relax
\mciteBstWouldAddEndPuncttrue
\mciteSetBstMidEndSepPunct{\mcitedefaultmidpunct}
{\mcitedefaultendpunct}{\mcitedefaultseppunct}\relax
\EndOfBibitem
\bibitem[Anisimov \latin{et~al.}(1997)Anisimov, Aryasetiawan, and
  Lichtenstein]{Anisimov.1997}
Anisimov,~V.~I.; Aryasetiawan,~F.; Lichtenstein,~A.~I. {First-principles
  calculations of the electronic structure and spectra of strongly correlated
  systems: the LDA+U method}. \emph{J. Phys. Condens. Matter} \textbf{1997},
  \emph{9}, 767\relax
\mciteBstWouldAddEndPuncttrue
\mciteSetBstMidEndSepPunct{\mcitedefaultmidpunct}
{\mcitedefaultendpunct}{\mcitedefaultseppunct}\relax
\EndOfBibitem
\bibitem[Dudarev \latin{et~al.}(1998)Dudarev, Botton, Savrasov, Humphreys, and
  Sutton]{Dudarev.1998}
Dudarev,~S.~L.; Botton,~G.~A.; Savrasov,~S.~Y.; Humphreys,~C.~J.; Sutton,~A.~P.
  {Electron-energy-loss spectra and the structural stability of nickel oxide:
  An LSDA+U study}. \emph{Phys. Rev. B} \textbf{1998}, \emph{57},
  1505--1509\relax
\mciteBstWouldAddEndPuncttrue
\mciteSetBstMidEndSepPunct{\mcitedefaultmidpunct}
{\mcitedefaultendpunct}{\mcitedefaultseppunct}\relax
\EndOfBibitem
\bibitem[Heyd \latin{et~al.}(2003)Heyd, Scuseria and Ernzerhof]{Heyd.2005}
Heyd, J.; Scuseria, G. E.; Ernzerhof, M. {Hybrid functionals based on a screened Coulomb potential.} \emph{J. Chem. Phys.} \textbf{2003}, \emph{118}, 8207--8215\relax
\mciteBstWouldAddEndPuncttrue
\mciteSetBstMidEndSepPunct{\mcitedefaultmidpunct}
{\mcitedefaultendpunct}{\mcitedefaultseppunct}\relax
\EndOfBibitem  
\bibitem[Rohrbach \latin{et~al.}(2004)Rohrbach]{Rohrbach.2004}
Rohrbach, A.; Hafner, J.; Kresse, G. {Ab initio study of the (0001) surfaces of hematite and chromia: Influence of strong electronic correlations.} \emph{Phys. Rev. B} \textbf{2024}, \emph{70}, 125426\relax
 \mciteBstWouldAddEndPuncttrue
\mciteSetBstMidEndSepPunct{\mcitedefaultmidpunct}
{\mcitedefaultendpunct}{\mcitedefaultseppunct}\relax
\EndOfBibitem  
\bibitem[Nolan \latin{et~al.}(2006)Nolan]{Nolan.2006}
Nolan, M.; Elliott, S. D. {The p-type conduction mechanism in Cu$_2$O: a first principles study.} \emph{Phys. Chem. Chem. Phys.} \textbf{2006}, \emph{8}, 5350--5358\relax
\mciteBstWouldAddEndPuncttrue
\mciteSetBstMidEndSepPunct{\mcitedefaultmidpunct}
{\mcitedefaultendpunct}{\mcitedefaultseppunct}\relax
\EndOfBibitem  
\bibitem[Lai \latin{et~al.}(2019)Lai, Song, Wan, Xue, Wang, Ye, Dai, Zhang,
  Yang, Du, and Yang]{Lai.2019}
Lai,~Y.; Song,~Z.; Wan,~Y.; Xue,~M.; Wang,~C.; Ye,~Y.; Dai,~L.; Zhang,~Z.;
  Yang,~W.; Du,~H. \latin{et~al.}  {Two-dimensional ferromagnetism and driven
  ferroelectricity in van der Waals CuCrP$_2$S$_6$}. \emph{Nanoscale}
  \textbf{2019}, \emph{11}, 5163--5170\relax
\mciteBstWouldAddEndPuncttrue
\mciteSetBstMidEndSepPunct{\mcitedefaultmidpunct}
{\mcitedefaultendpunct}{\mcitedefaultseppunct}\relax
\EndOfBibitem
\bibitem[Studenyak \latin{et~al.}(2003)Studenyak, Mykajlo, Vysochanskii, and
  Cajipe]{Studenyak.2003}
Studenyak,~I.~P.; Mykajlo,~O.~A.; Vysochanskii,~Y.~M.; Cajipe,~V.~B. {Optical
  absorption studies of phase transitions in CuCrP$_2$S$_6$ layered
  antiferroelectrics}. \emph{J. Phys.: Condens. Matter} \textbf{2003},
  \emph{15}, 6773\relax
\mciteBstWouldAddEndPuncttrue
\mciteSetBstMidEndSepPunct{\mcitedefaultmidpunct}
{\mcitedefaultendpunct}{\mcitedefaultseppunct}\relax
\EndOfBibitem
\bibitem[Io \latin{et~al.}(2023)Io, Pang, Wong, Zhao, Ding, Mao, Zhao, Guo,
  Yuan, Zhao, Yi, and Hao]{Io.2023}
Io,~W.~F.; Pang,~S.~Y.; Wong,~L.~W.; Zhao,~Y.; Ding,~R.; Mao,~J.; Zhao,~Y.;
  Guo,~F.; Yuan,~S.; Zhao,~J. \latin{et~al.}  {Direct observation of intrinsic
  room-temperature ferroelectricity in 2D layered CuCrP$_2$S$_6$}. \emph{Nat.
  Commun.} \textbf{2023}, \emph{14}, 7304\relax
\mciteBstWouldAddEndPuncttrue
\mciteSetBstMidEndSepPunct{\mcitedefaultmidpunct}
{\mcitedefaultendpunct}{\mcitedefaultseppunct}\relax
\EndOfBibitem
\bibitem[Rao and Susner(2023)Rao, and Susner]{Rao.2023}
Rao,~R.; Susner,~M.~A. {Phonon anharmonicity in Cu-based layered
  thiophosphates}. \emph{Mater. Today Commun.} \textbf{2023}, \emph{35},
  105840\relax
\mciteBstWouldAddEndPuncttrue
\mciteSetBstMidEndSepPunct{\mcitedefaultmidpunct}
{\mcitedefaultendpunct}{\mcitedefaultseppunct}\relax
\EndOfBibitem
\bibitem[Ma \latin{et~al.}(2023)Ma, Yan, Luo, Pazos, Zhang, Lv, Chen, Liu,
  Wang, Chen, Li, Zheng, Lin, Algaidi, Sun, Liu, Tu, Alshareef, Gong, Lanza,
  Xue, and Zhang]{Ma.2023}
Ma,~Y.; Yan,~Y.; Luo,~L.; Pazos,~S.; Zhang,~C.; Lv,~X.; Chen,~M.; Liu,~C.;
  Wang,~Y.; Chen,~A. \latin{et~al.}  {High-performance van der Waals
  antiferroelectric CuCrP$_2$S$_6$-based memristors}. \emph{Nat. Commun.}
  \textbf{2023}, \emph{14}, 7891\relax
\mciteBstWouldAddEndPuncttrue
\mciteSetBstMidEndSepPunct{\mcitedefaultmidpunct}
{\mcitedefaultendpunct}{\mcitedefaultseppunct}\relax
\EndOfBibitem
\bibitem[Wang \latin{et~al.}(2018)Wang, Li, Zhu, Boscoboinik, and
  Zhou]{Wang.2018oja}
Wang,~J.; Li,~C.; Zhu,~Y.; Boscoboinik,~J.~A.; Zhou,~G. {Insight into the phase
  transformation pathways of copper oxidation: From oxygen chemisorption on the
  clean surface to multilayer bulk oxide growth}. \emph{J. Phys. Chem. C}
  \textbf{2018}, \emph{122}, 26519--26527\relax
\mciteBstWouldAddEndPuncttrue
\mciteSetBstMidEndSepPunct{\mcitedefaultmidpunct}
{\mcitedefaultendpunct}{\mcitedefaultseppunct}\relax
\EndOfBibitem
\bibitem[Liu \latin{et~al.}(2022)Liu, Huber, Spronsen, Salmeron, and
  Bluhm]{Liu.2022}
Liu,~B.-H.; Huber,~M.; Spronsen,~M. A.~v.; Salmeron,~M.; Bluhm,~H. {Ambient
  pressure X-ray photoelectron spectroscopy study of room-temperature oxygen
  adsorption on Cu(100) and Cu(111)}. \emph{Appl. Surf. Sci.} \textbf{2022},
  \emph{583}, 152438\relax
\mciteBstWouldAddEndPuncttrue
\mciteSetBstMidEndSepPunct{\mcitedefaultmidpunct}
{\mcitedefaultendpunct}{\mcitedefaultseppunct}\relax
\EndOfBibitem
\bibitem[Chambers and Droubay(2001)Chambers, and Droubay]{Chambers.2001}
Chambers,~S.~A.; Droubay,~T. {Role of oxide ionicity in electronic screening at
  oxide/metal interfaces}. \emph{Phys. Rev. B} \textbf{2001}, \emph{64},
  075410\relax
\mciteBstWouldAddEndPuncttrue
\mciteSetBstMidEndSepPunct{\mcitedefaultmidpunct}
{\mcitedefaultendpunct}{\mcitedefaultseppunct}\relax
\EndOfBibitem
\bibitem[Biesinger \latin{et~al.}(2004)Biesinger, Brown, Mycroft, Davidson, and
  McIntyre]{Biesinger.2004}
Biesinger,~M.~C.; Brown,~C.; Mycroft,~J.~R.; Davidson,~R.~D.; McIntyre,~N.~S.
  {X‐ray photoelectron spectroscopy studies of chromium compounds}.
  \emph{Surf. Interface Anal.} \textbf{2004}, \emph{36}, 1550--1563\relax
\mciteBstWouldAddEndPuncttrue
\mciteSetBstMidEndSepPunct{\mcitedefaultmidpunct}
{\mcitedefaultendpunct}{\mcitedefaultseppunct}\relax
\EndOfBibitem
\bibitem[Bataillou \latin{et~al.}(2020)Bataillou, Martinelli, Desgranges,
  Bosonnet, Ginestar, Miserque, Wouters, Latu-Romain, Pugliara, Proietti, and
  Monceau]{Bataillou.2020}
Bataillou,~L.; Martinelli,~L.; Desgranges,~C.; Bosonnet,~S.; Ginestar,~K.;
  Miserque,~F.; Wouters,~Y.; Latu-Romain,~L.; Pugliara,~A.; Proietti,~A.
  \latin{et~al.}  {Growth kinetics and characterization of chromia scales
  formed on Ni-30Cr alloy in impure argon at $700^\circ$C}. \emph{Oxid. Met.}
  \textbf{2020}, \emph{93}, 329--353\relax
\mciteBstWouldAddEndPuncttrue
\mciteSetBstMidEndSepPunct{\mcitedefaultmidpunct}
{\mcitedefaultendpunct}{\mcitedefaultseppunct}\relax
\EndOfBibitem
\bibitem[Pinho \latin{et~al.}(2021)Pinho, Chartier, Miserque, Menut, and
  Moussy]{Pinho.2021}
Pinho,~P. V.~B.; Chartier,~A.; Miserque,~F.; Menut,~D.; Moussy,~J.~B. {Impact
  of epitaxial strain on crystal field splitting of $\alpha$-Cr$_2$O$_3$(0001) thin films
  quantified by X-ray photoemission spectroscopy}. \emph{Mater. Res. Lett.}
  \textbf{2021}, \emph{9}, 163--168\relax
\mciteBstWouldAddEndPuncttrue
\mciteSetBstMidEndSepPunct{\mcitedefaultmidpunct}
{\mcitedefaultendpunct}{\mcitedefaultseppunct}\relax
\EndOfBibitem
\bibitem[Theil \latin{et~al.}(1998)Theil, Elp, and Folkmann]{Theil.1998}
Theil,~C.; Elp,~J.~v.; Folkmann,~F. {Ligand field parameters obtained from and
  chemical shifts observed at the Cr $L_{2,3}$ edges}. \emph{Phys. Rev. B}
  \textbf{1998}, \emph{59}, 7931--7936\relax
\mciteBstWouldAddEndPuncttrue
\mciteSetBstMidEndSepPunct{\mcitedefaultmidpunct}
{\mcitedefaultendpunct}{\mcitedefaultseppunct}\relax
\EndOfBibitem
\bibitem[Dedkov \latin{et~al.}(2005)Dedkov, Vinogradov, Fonin, König, Vyalikh,
  Preobrajenski, Krasnikov, Kleimenov, Nesterov, Rüdiger, Molodtsov, and
  Güntherodt]{Dedkov.2005}
Dedkov,~Y.~S.; Vinogradov,~A.~S.; Fonin,~M.; König,~C.; Vyalikh,~D.~V.;
  Preobrajenski,~A.~B.; Krasnikov,~S.~A.; Kleimenov,~E.~Y.; Nesterov,~M.~A.;
  Rüdiger,~U. \latin{et~al.}  {Correlations in the electronic structure of
  half-metallic ferromagnetic CrO$_2$ films: An x-ray absorption and resonant
  photoemission spectroscopy study}. \emph{Phys. Rev. B} \textbf{2005},
  \emph{72}, 060401\relax
\mciteBstWouldAddEndPuncttrue
\mciteSetBstMidEndSepPunct{\mcitedefaultmidpunct}
{\mcitedefaultendpunct}{\mcitedefaultseppunct}\relax
\EndOfBibitem
\bibitem[Brik \latin{et~al.}(2004)Brik, Avram, and Avram]{Brik.2004}
Brik,~M.; Avram,~N.; Avram,~C. {Crystal field analysis of energy level
  structure of the Cr$_2$O$_3$ antiferromagnet}. \emph{Solid State Commun.}
  \textbf{2004}, \emph{132}, 831--835\relax
\mciteBstWouldAddEndPuncttrue
\mciteSetBstMidEndSepPunct{\mcitedefaultmidpunct}
{\mcitedefaultendpunct}{\mcitedefaultseppunct}\relax
\EndOfBibitem
\bibitem[Grioni \latin{et~al.}(1991)Grioni, Acker, Czyžyk, and
  Fuggle]{Grioni.1991}
Grioni,~M.; Acker,~J. F.~v.; Czyžyk,~M.~T.; Fuggle,~J.~C. {Unoccupied
  electronic structure and core-hole effects in the x-ray-absorption spectra of
  Cu$_2$O}. \emph{Phys. Rev. B} \textbf{1991}, \emph{45}, 3309--3318\relax
\mciteBstWouldAddEndPuncttrue
\mciteSetBstMidEndSepPunct{\mcitedefaultmidpunct}
{\mcitedefaultendpunct}{\mcitedefaultseppunct}\relax
\EndOfBibitem
\bibitem[Vegelius \latin{et~al.}(2012)Vegelius, Kvashnina, Hollmark,
  Klintenberg, Kvashnin, Soroka, Werme, and Butorin]{Vegelius.2012}
Vegelius,~J.~R.; Kvashnina,~K.~O.; Hollmark,~H.; Klintenberg,~M.;
  Kvashnin,~Y.~O.; Soroka,~I.~L.; Werme,~L.; Butorin,~S.~M. {X‑ray
  spectroscopic study of Cu$_2$S, CuS, and copper films exposed to Na$_2$S
  solutions}. \emph{J. Phys. Chem. C} \textbf{2012}, \emph{116},
  22293--22300\relax
\mciteBstWouldAddEndPuncttrue
\mciteSetBstMidEndSepPunct{\mcitedefaultmidpunct}
{\mcitedefaultendpunct}{\mcitedefaultseppunct}\relax
\EndOfBibitem
\bibitem[Jiang \latin{et~al.}(2013)Jiang, Prendergast, Borondics, Porsgaard,
  Giovanetti, Pach, Newberg, Bluhm, Besenbacher, and Salmeron]{Jiang.2013}
Jiang,~P.; Prendergast,~D.; Borondics,~F.; Porsgaard,~S.; Giovanetti,~L.;
  Pach,~E.; Newberg,~J.; Bluhm,~H.; Besenbacher,~F.; Salmeron,~M. {Experimental
  and theoretical investigation of the electronic structure of Cu$_2$O and CuO
  thin films on Cu(110) using x-ray photoelectron and absorption spectroscopy}.
  \emph{J. Chem. Phys.} \textbf{2013}, \emph{138}, 024704\relax
\mciteBstWouldAddEndPuncttrue
\mciteSetBstMidEndSepPunct{\mcitedefaultmidpunct}
{\mcitedefaultendpunct}{\mcitedefaultseppunct}\relax
\EndOfBibitem
\bibitem[Henzler \latin{et~al.}(2015)Henzler, Heilemann, Kneer, Guttmann, Jia,
  Bartsch, Lu, and Palzer]{Henzler.2015}
Henzler,~K.; Heilemann,~A.; Kneer,~J.; Guttmann,~P.; Jia,~H.; Bartsch,~E.;
  Lu,~Y.; Palzer,~S. {Investigation of reactions between trace gases and
  functional CuO nanospheres and octahedrons using NEXAFS-TXM imaging}.
  \emph{Sci. Rep.} \textbf{2015}, \emph{5}, 17729\relax
\mciteBstWouldAddEndPuncttrue
\mciteSetBstMidEndSepPunct{\mcitedefaultmidpunct}
{\mcitedefaultendpunct}{\mcitedefaultseppunct}\relax
\EndOfBibitem
\bibitem[Leapman \latin{et~al.}(1982)Leapman, Grunes, and Fejes]{Leapman.1982}
Leapman,~R.~D.; Grunes,~L.~A.; Fejes,~P.~L. {Study of the $L_{2,3}$ edges in the 3d
  transition metals and their oxides by electron-energy-loss spectroscopy with
  comparisons to theory}. \emph{Phys. Rev. B} \textbf{1982}, \emph{26},
  614--635\relax
\mciteBstWouldAddEndPuncttrue
\mciteSetBstMidEndSepPunct{\mcitedefaultmidpunct}
{\mcitedefaultendpunct}{\mcitedefaultseppunct}\relax
\EndOfBibitem
\bibitem[Muller \latin{et~al.}(1982)Muller, Jepsen, and Wilkins]{Muller.1982}
Muller,~J.; Jepsen,~O.; Wilkins,~J. {X-ray absorption spectra: $K$-edges of 3d
  transition metals, $L$-edges of 3d and 4d metals, and $M$-edges of palladium}.
  \emph{Solid State Commun.} \textbf{1982}, \emph{42}, 365--368\relax
\mciteBstWouldAddEndPuncttrue
\mciteSetBstMidEndSepPunct{\mcitedefaultmidpunct}
{\mcitedefaultendpunct}{\mcitedefaultseppunct}\relax
\EndOfBibitem
\bibitem[Heinemann \latin{et~al.}(2013)Heinemann, Eifert, and
  Heiliger]{Heinemann.2013}
Heinemann,~M.; Eifert,~B.; Heiliger,~C. {Band structure and phase stability of
  the copper oxides Cu$_2$O, CuO, and Cu$_4$O$_3$}. \emph{Phys. Rev. B} \textbf{2013},
  \emph{87}, 115111\relax
\mciteBstWouldAddEndPuncttrue
\mciteSetBstMidEndSepPunct{\mcitedefaultmidpunct}
{\mcitedefaultendpunct}{\mcitedefaultseppunct}\relax
\EndOfBibitem
\bibitem[Visibile \latin{et~al.}(2019)Visibile, Wang, Vertova, Rondinini,
  Minguzzi, Ahlberg, and Busch]{Visibile.2019}
Visibile,~A.; Wang,~R.~B.; Vertova,~A.; Rondinini,~S.; Minguzzi,~A.;
  Ahlberg,~E.; Busch,~M. {Influence of strain on the band gap of Cu$_2$O}.
  \emph{Chem. Mater.} \textbf{2019}, \emph{31}, 4787--4792\relax
\mciteBstWouldAddEndPuncttrue
\mciteSetBstMidEndSepPunct{\mcitedefaultmidpunct}
{\mcitedefaultendpunct}{\mcitedefaultseppunct}\relax
\EndOfBibitem
\bibitem[Zaanen \latin{et~al.}(1985)Zaanen, Sawatzky, and Allen]{Zaanen.1985}
Zaanen,~J.; Sawatzky,~G.~A.; Allen,~J.~W. {Band gaps and electronic structure
  of transition-metal compounds}. \emph{Phys. Rev. Lett.} \textbf{1985},
  \emph{55}, 418--421\relax
\mciteBstWouldAddEndPuncttrue
\mciteSetBstMidEndSepPunct{\mcitedefaultmidpunct}
{\mcitedefaultendpunct}{\mcitedefaultseppunct}\relax
\EndOfBibitem
\end{mcitethebibliography}

\begin{mcitethebibliography}{29}
\providecommand*\natexlab[1]{#1}
\providecommand*\mciteSetBstSublistMode[1]{}
\providecommand*\mciteSetBstMaxWidthForm[2]{}
\providecommand*\mciteBstWouldAddEndPuncttrue
  {\def\EndOfBibitem{\unskip.}}
\providecommand*\mciteBstWouldAddEndPunctfalse
  {\let\EndOfBibitem\relax}
\providecommand*\mciteSetBstMidEndSepPunct[3]{}
\providecommand*\mciteSetBstSublistLabelBeginEnd[3]{}
\providecommand*\EndOfBibitem{}
\mciteSetBstSublistMode{f}
\mciteSetBstMaxWidthForm{subitem}{(\alph{mcitesubitemcount})}
\mciteSetBstSublistLabelBeginEnd
  {\mcitemaxwidthsubitemform\space}
  {\relax}
  {\relax}
 \bibitem[Maisonneuve \latin{et~al.}(1993)Maisonneuve, Cajipe, and
  Payen]{Maisonneuve.1993abc}
Maisonneuve,~V.; Cajipe,~V.~B.; Payen,~C. {Low-temperature neutron powder
  diffraction study of copper chromium thiophosphate (CuCrP$_2$S$_6$): observation of
  an ordered, antipolar copper sublattice}. \emph{Chem. Mater.} \textbf{1993},
  \emph{5}, 758--760\relax
\mciteBstWouldAddEndPuncttrue
\mciteSetBstMidEndSepPunct{\mcitedefaultmidpunct}
{\mcitedefaultendpunct}{\mcitedefaultseppunct}\relax
\EndOfBibitem
\bibitem[Colombet \latin{et~al.}(1982)Colombet, Leblanc, Danot, and
  Rouxel]{Colombet.1982}
Colombet,~P.; Leblanc,~A.; Danot,~M.; Rouxel,~J. {Structural aspects and
  magnetic properties of the lamellar compound Cu$_{0.50}$Cr$_{0.50}$PS$_3$}. \emph{J. Solid
  State Chem.} \textbf{1982}, \emph{41}, 174--184\relax
\mciteBstWouldAddEndPuncttrue
\mciteSetBstMidEndSepPunct{\mcitedefaultmidpunct}
{\mcitedefaultendpunct}{\mcitedefaultseppunct}\relax
\EndOfBibitem
\bibitem[Wang \latin{et~al.}(2018)Wang, Li, Zhu, Boscoboinik, and
  Zhou]{Wang.2018oja}
Wang,~J.; Li,~C.; Zhu,~Y.; Boscoboinik,~J.~A.; Zhou,~G. {Insight into the phase
  transformation pathways of copper oxidation: From oxygen chemisorption on the
  clean surface to multilayer bulk oxide growth}. \emph{J. Phys. Chem. C}
  \textbf{2018}, \emph{122}, 26519--26527\relax
\mciteBstWouldAddEndPuncttrue
\mciteSetBstMidEndSepPunct{\mcitedefaultmidpunct}
{\mcitedefaultendpunct}{\mcitedefaultseppunct}\relax
\EndOfBibitem
\bibitem[Liu \latin{et~al.}(2022)Liu, Huber, Spronsen, Salmeron, and
  Bluhm]{Liu.2022}
Liu,~B.-H.; Huber,~M.; Spronsen,~M. A.~v.; Salmeron,~M.; Bluhm,~H. {Ambient
  pressure X-ray photoelectron spectroscopy study of room-temperature oxygen
  adsorption on Cu(100) and Cu(111)}. \emph{Appl. Surf. Sci.} \textbf{2022},
  \emph{583}, 152438\relax
\mciteBstWouldAddEndPuncttrue
\mciteSetBstMidEndSepPunct{\mcitedefaultmidpunct}
{\mcitedefaultendpunct}{\mcitedefaultseppunct}\relax
\EndOfBibitem
\bibitem[Hou \latin{et~al.}(2020)Hou, Zhang, Ma, Tang, Hao, Cheng, and
  Qiu]{Hou.2020}
Hou,~X.; Zhang,~X.; Ma,~Q.; Tang,~X.; Hao,~Q.; Cheng,~Y.; Qiu,~T. {Alloy
  engineering in few‐layer manganese phosphorus trichalcogenides for
  surface‐enhanced raman scattering}. \emph{Adv. Funct. Mater.}
  \textbf{2020}, \emph{30}, 1910171\relax
\mciteBstWouldAddEndPuncttrue
\mciteSetBstMidEndSepPunct{\mcitedefaultmidpunct}
{\mcitedefaultendpunct}{\mcitedefaultseppunct}\relax
\EndOfBibitem
\bibitem[Jin \latin{et~al.}(2022)Jin, Yan, Jin, Li, Dedkov, and
  Voloshina]{Jin.2022yb}
Jin,~Y.; Yan,~M.; Jin,~Y.; Li,~K.; Dedkov,~Y.; Voloshina,~E. {To the stability
  of Janus phases in layered trichalcogenide MPX$_3$ crystals: Insights from
  experiments and theory}. \emph{J. Phys. Chem. C} \textbf{2022}, \emph{126},
  16061--16068\relax
\mciteBstWouldAddEndPuncttrue
\mciteSetBstMidEndSepPunct{\mcitedefaultmidpunct}
{\mcitedefaultendpunct}{\mcitedefaultseppunct}\relax
\EndOfBibitem
\bibitem[Kim \latin{et~al.}(2021)Kim, Lee, Lee, and Ryu]{Kim.2021asd}
Kim,~S.; Lee,~J.; Lee,~C.; Ryu,~S. {Polarized Raman spectra and complex Raman
  tensors of antiferromagnetic semiconductor CrPS$_4$}. \emph{J. Phys. Chem. C}
  \textbf{2021}, \emph{125}, 2691--2698\relax
\mciteBstWouldAddEndPuncttrue
\mciteSetBstMidEndSepPunct{\mcitedefaultmidpunct}
{\mcitedefaultendpunct}{\mcitedefaultseppunct}\relax
\EndOfBibitem
\bibitem[Susner \latin{et~al.}(2020)Susner, Rao, Pelton, McLeod, and
  Maruyama]{Susner.2020}
Susner,~M.~A.; Rao,~R.; Pelton,~A.~T.; McLeod,~M.~V.; Maruyama,~B.
  {Temperature-dependent Raman scattering and x-ray diffraction study of phase
  transitions in layered multiferroic CuCrP$_2$S$_6$}. \emph{Phys. Rev. Mater.}
  \textbf{2020}, \emph{4}, 104003\relax
\mciteBstWouldAddEndPuncttrue
\mciteSetBstMidEndSepPunct{\mcitedefaultmidpunct}
{\mcitedefaultendpunct}{\mcitedefaultseppunct}\relax
\EndOfBibitem
\bibitem[Io \latin{et~al.}(2023)Io, Pang, Wong, Zhao, Ding, Mao, Zhao, Guo,
  Yuan, Zhao, Yi, and Hao]{Io.2023}
Io,~W.~F.; Pang,~S.~Y.; Wong,~L.~W.; Zhao,~Y.; Ding,~R.; Mao,~J.; Zhao,~Y.;
  Guo,~F.; Yuan,~S.; Zhao,~J.; Yi,~J.; Hao,~J. {Direct observation of intrinsic
  room-temperature ferroelectricity in 2D layered CuCrP$_2$S$_6$}. \emph{Nat.
  Commun.} \textbf{2023}, \emph{14}, 7304\relax
\mciteBstWouldAddEndPuncttrue
\mciteSetBstMidEndSepPunct{\mcitedefaultmidpunct}
{\mcitedefaultendpunct}{\mcitedefaultseppunct}\relax
\EndOfBibitem
\bibitem[Rao and Susner(2023)Rao, and Susner]{Rao.2023}
Rao,~R.; Susner,~M.~A. {Phonon anharmonicity in Cu-based layered
  thiophosphates}. \emph{Mater. Today Commun.} \textbf{2023}, \emph{35},
  105840\relax
\mciteBstWouldAddEndPuncttrue
\mciteSetBstMidEndSepPunct{\mcitedefaultmidpunct}
{\mcitedefaultendpunct}{\mcitedefaultseppunct}\relax
\EndOfBibitem
\bibitem[Ma \latin{et~al.}(2023)Ma, Yan, Luo, Pazos, Zhang, Lv, Chen, Liu,
  Wang, Chen, Li, Zheng, Lin, Algaidi, Sun, Liu, Tu, Alshareef, Gong, Lanza,
  Xue, and Zhang]{Ma.2023}
Ma,~Y.; Yan,~Y.; Luo,~L.; Pazos,~S.; Zhang,~C.; Lv,~X.; Chen,~M.; Liu,~C.;
  Wang,~Y.; Chen,~A.; Li,~Y.; Zheng,~D.; Lin,~R.; Algaidi,~H.; Sun,~M.;
  Liu,~J.~Z.; Tu,~S.; Alshareef,~H.~N.; Gong,~C.; Lanza,~M. \latin{et~al.}
  {High-performance van der Waals antiferroelectric CuCrP$_2$S$_6$-based memristors}.
  \emph{Nat. Commun.} \textbf{2023}, \emph{14}, 7891\relax
\mciteBstWouldAddEndPuncttrue
\mciteSetBstMidEndSepPunct{\mcitedefaultmidpunct}
{\mcitedefaultendpunct}{\mcitedefaultseppunct}\relax
\EndOfBibitem
\bibitem[Lee \latin{et~al.}(2017)Lee, Ko, Kim, Bark, Kang, Jung, Park, Lee,
  Ryu, and Lee]{Lee.2017f9o}
Lee,~J.; Ko,~T.~Y.; Kim,~J.~H.; Bark,~H.; Kang,~B.; Jung,~S.-G.; Park,~T.;
  Lee,~Z.; Ryu,~S.; Lee,~C. {Structural and optical properties of single- and
  few-layer magnetic semiconductor CrPS$_4$}. \emph{ACS Nano} \textbf{2017},
  \emph{11}, 10935--10944\relax
\mciteBstWouldAddEndPuncttrue
\mciteSetBstMidEndSepPunct{\mcitedefaultmidpunct}
{\mcitedefaultendpunct}{\mcitedefaultseppunct}\relax
\EndOfBibitem
\bibitem[Wu and Chen(2019)Wu, and Chen]{Wu.2019wly}
Wu,~H.; Chen,~H. {Probing the properties of lattice vibrations and surface
  electronic states in magnetic semiconductor CrPS$_4$}. \emph{RSC Adv.}
  \textbf{2019}, \emph{9}, 30655--30658\relax
\mciteBstWouldAddEndPuncttrue
\mciteSetBstMidEndSepPunct{\mcitedefaultmidpunct}
{\mcitedefaultendpunct}{\mcitedefaultseppunct}\relax
\EndOfBibitem
\end{mcitethebibliography}

\clearpage
\begin{table}[]
    \centering
\setlength{\tabcolsep}{21pt} 
\begin{tabular}{ccc}
\hline
Parameter & DFT &  XRD \\ \hline
$a$    & 5.952   &  5.92   \\
$b$    & 10.309   &  10.247    \\
$c$     & 13.190   &  12.974   \\
$\alpha=\gamma$     & 90   &  90   \\
$\beta$     & 98.65   &  99.28   \\  \hline
\end{tabular}
    \caption{Lattice parameters of the CuCrP$_2$S$_6$ bulk from DFT calculations and XRD experiments.}
    \label{tab:lattice_parameters}
\end{table}

\clearpage
\begin{table}[]
    \centering
\setlength{\tabcolsep}{21pt} 
\begin{tabular}{ccc}
\hline
Element & SEM/EDX &  TEM/EDX \\ \hline
Cu    & 12.26   &  16.02   \\
Cr    & 10.07   &  9.67    \\
P     & 19.59   &  19.37   \\
S     & 58.08   &  54.94   \\  \hline
\end{tabular}
    \caption{Atomic concentrations of elements in synthesised CuCrP$_2$S$_6$ bulk crystals obtained in SEM/EDX and TEM/EDX measurements.}
    \label{tab:SEM_EDX_TEM_EDX}
\end{table}

\clearpage
\begin{table}[]
    \centering
\footnotesize
\begin{tabular}{l ccccc}
\hline
Magn. state & $E_\mathrm{tot}$ & $m_\mathrm{Cr}$& $m_\mathrm{Cu}$& $m_\mathrm{P}$& $m_\mathrm{S}$ \\ 
\hline
\multicolumn{6}{l}{1ML-CCPS:}\\
FM	&$-200.5386$	&$+3.411$ 	&$-0.012$		&$+0.011$	&$-0.066$\\
AFM	&$-200.4875$	&$\pm3.396$	&$\pm0.015$	&$\pm0.003$	&$\pm0.052$\\[6pt]
\multicolumn{6}{l}{bulk CCPS:}\\
FM	&$-1211.6706$ &$+3.402^a$/$+3.402^b$	&$-0.010^a$/$-0.010^b$	&$+0.011^a$/$+0.011^b$	&$-0.063^a$/$-0.063^b$	    \\
AFM	&$-1211.6772$	&$+3.402^a$/$-3.402^b$	&$-0.010^a$/$+0.010^b$	&$+0.011^a$/$-0.011^b$	&$-0.063^a$/$+0.063^b$	    \\
 \hline
\end{tabular}

$^a$ average magnetic moment per layer A (see Fig.\,1 of the main text)\\
$^b$ average magnetic moment per layer B (see Fig.\,1 of the main text)

    \caption{Results obtained for the different magnetic states of 1ML-CuCrP$_2$S$_6$ (1ML-CCPS) and bulk CuCrP$_2$S$_6$ (bulk CCPS) with PBE$+U+$D2: $E_\mathrm{tot}$ (in eV per hexagonal unit cell) is the total energy;  magnetic moments of Cr, Cu, P, and S  ($m$, in $\mu_B$). }
    \label{tab:Magnetic}
    
\end{table}

\clearpage
\begin{figure}
\includegraphics[width=0.75\textwidth]{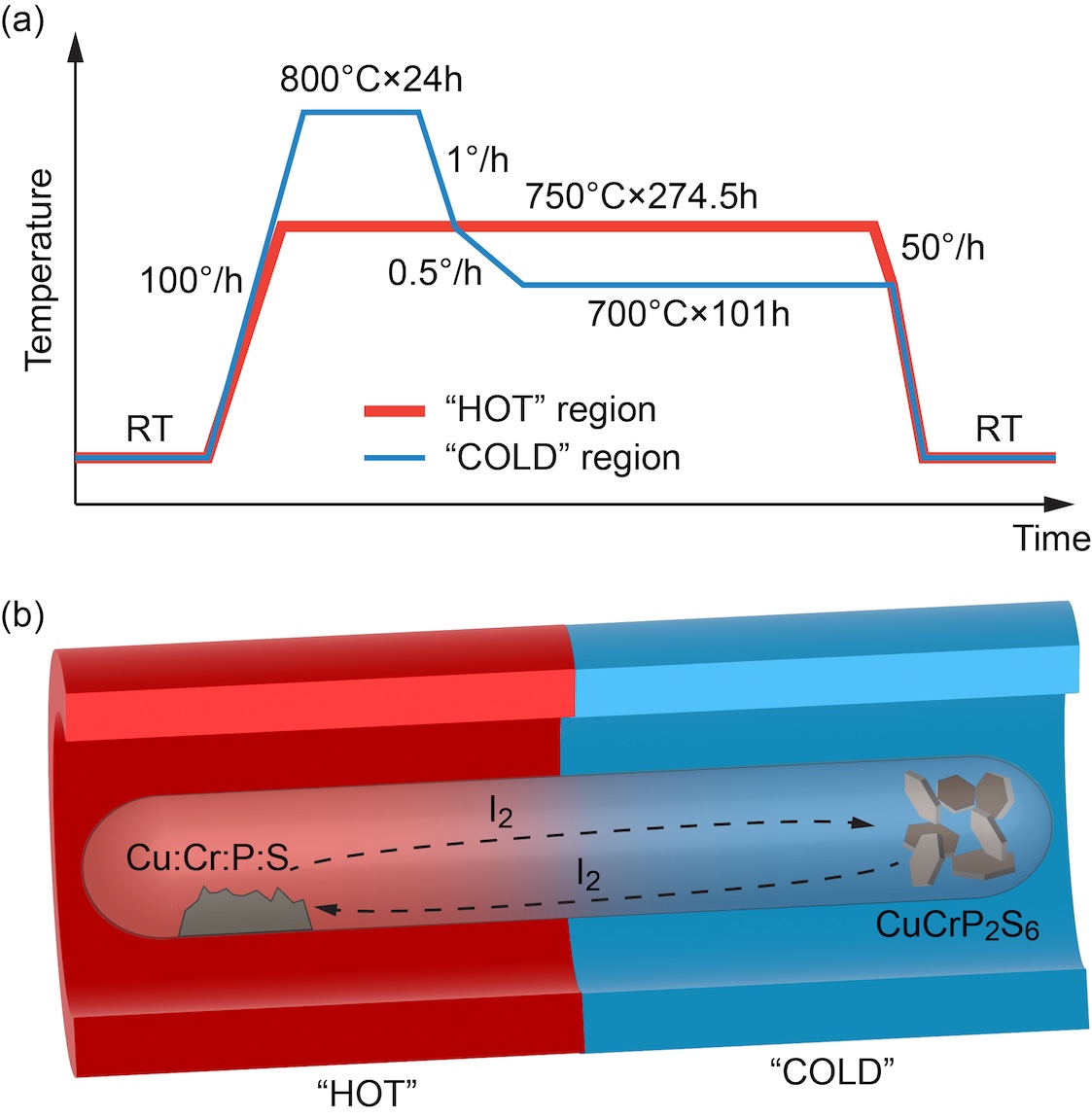}\\
Fig.\,S1. Synthesis scheme used in the present work: (a) Temperature as a function of time profile used for the CVT synthesis of CuCrP$_2$S$_6$; (b) Schematic drawing of an ampule during CVT growth with arrows indicating the mass ﬂow.
\end{figure}

\clearpage
\begin{figure}
\includegraphics[width=\textwidth]{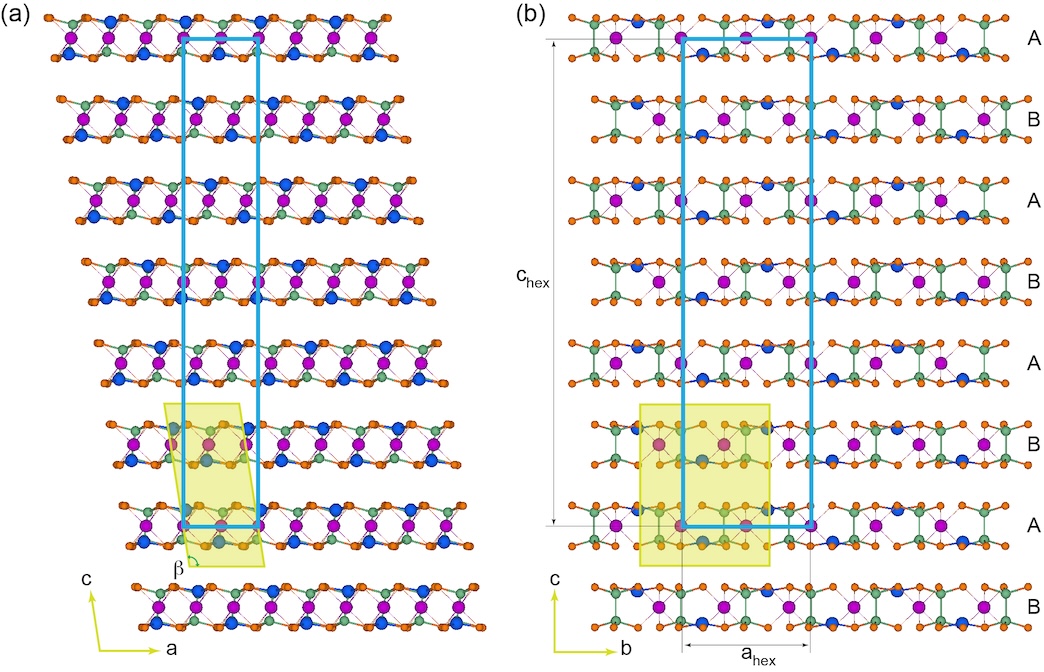}\\
Fig.\,S2. Crystallographic structure of CuCrP$_2$S$_6$ in two representations used in the present group: (a) $C2/c$ space group and (b) hexagonal unit cell containing $6$ layers.
\end{figure}

\clearpage
\begin{figure}
\includegraphics[width=1.0\textwidth]{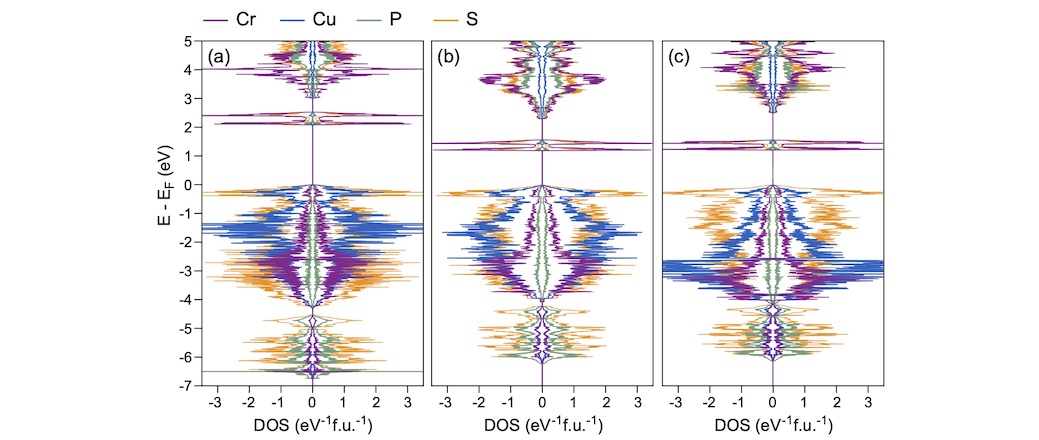}\\
Fig.\,S3. DOS plots for bulk CuCrP$_2$S$_6$ AFM ground state obtained with (a) HSE06 functional, (b) PBE$+U$ with $U(\textrm{Cr})=4$\,eV and  $U(\textrm{Cu})=4$\,eV, and (c) PBE$+U$ with $U(\textrm{Cr})=5$\,eV and  $U(\textrm{Cu})=7$\,eV.
\end{figure}

\clearpage
\begin{figure}
\includegraphics[width=0.75\textwidth]{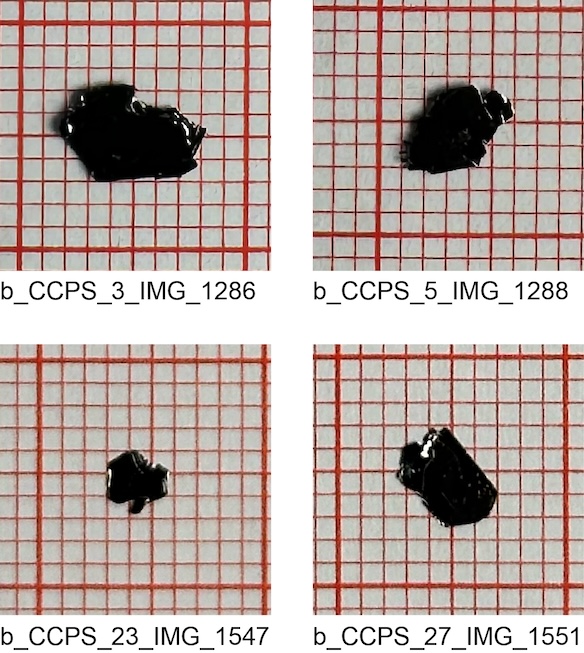}\\
Fig.\,S4. Representative photos of several CuCrP$_2$S$_6$ bulk crystals used in the present work.
\end{figure}

\clearpage
\begin{figure}
\includegraphics[width=0.4\textwidth]{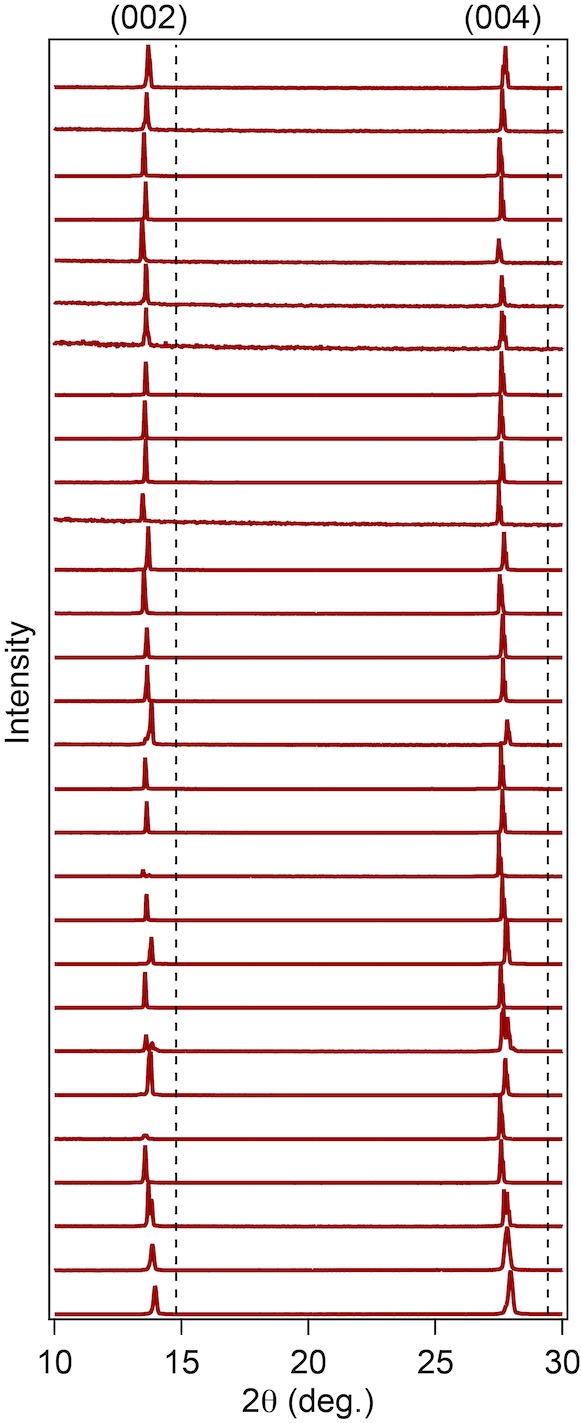}\\
Fig.\,S5. Series of XRD patterns measured over $30$ CuCrP$_2$S$_6$ crystals in the $2\theta$ range corresponding to $(002)$ and $(004)$ diffraction peaks. Vertical dashed lines mark the positions of the respective diffraction peaks for the CrPS$_4$ crystal.
\end{figure}

\clearpage
\begin{figure}
\includegraphics[width=\textwidth]{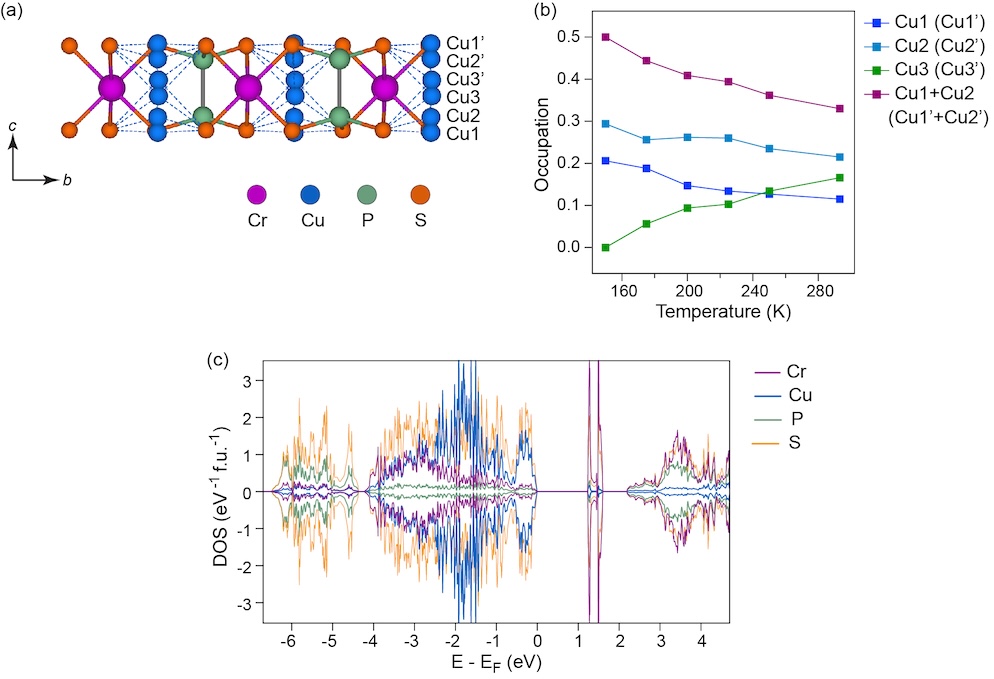}\\
Fig.\,S6. (a) Crystallographic structure of CuCrP$_2$S$_6$ obtained in the structure determination. Cu1 (Cu1'), Cu2 (Cu2'), and Cu3 (Cu3') denote six Cu positions which occupations were determined at different temperatures. (b) Occupation numbers for Cu1 (Cu1'), Cu2 (Cu2'), and Cu3 (Cu3'). (c) DOS calculated using the crystallographic structure of CuCrP$_2$S$_6$ bulk obtained in the refinement procedure.
\end{figure}

\clearpage
\begin{figure}
\includegraphics[width=0.75\textwidth]{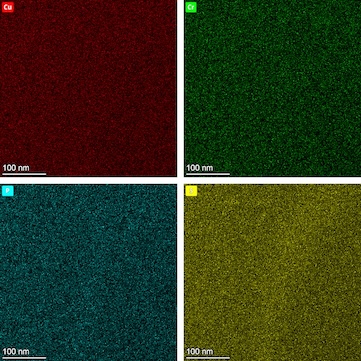}\\
Fig.\,S7. EDX elements' distribution maps obtained in TEM measurements of CuCrP$_2$S$_6$ bulk crystal.
\end{figure}

\clearpage
\begin{figure}
\includegraphics[width=0.75\textwidth]{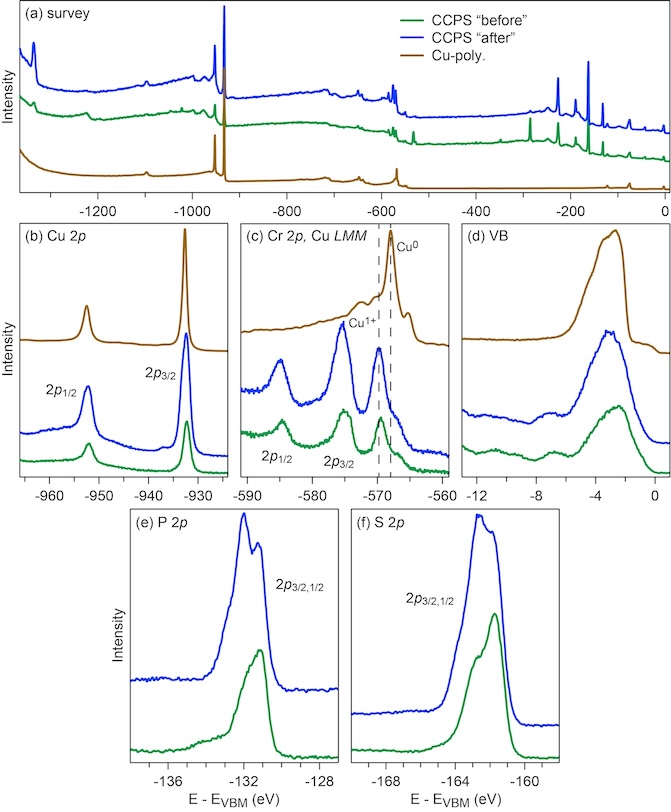}\\
Fig.\,S8. XPS spectra of CuCrP$_2$S$_6$ (CCPS) bulk crystal (before and after cleavage) and Cu-poly sample collected under laboratory conditions: (a) survey, (b) Cu\,$2p$, (c) Cr\,$2p$ and Cu\,$LMM$, (d) valence band, (e) P\,$2p$, (f) S\,$2p$. All spectra were collected with photon energy $h\nu=1486.6$\,eV (Al\,$K\alpha$).
\end{figure}

\clearpage
\begin{figure}
\includegraphics[width=\textwidth]{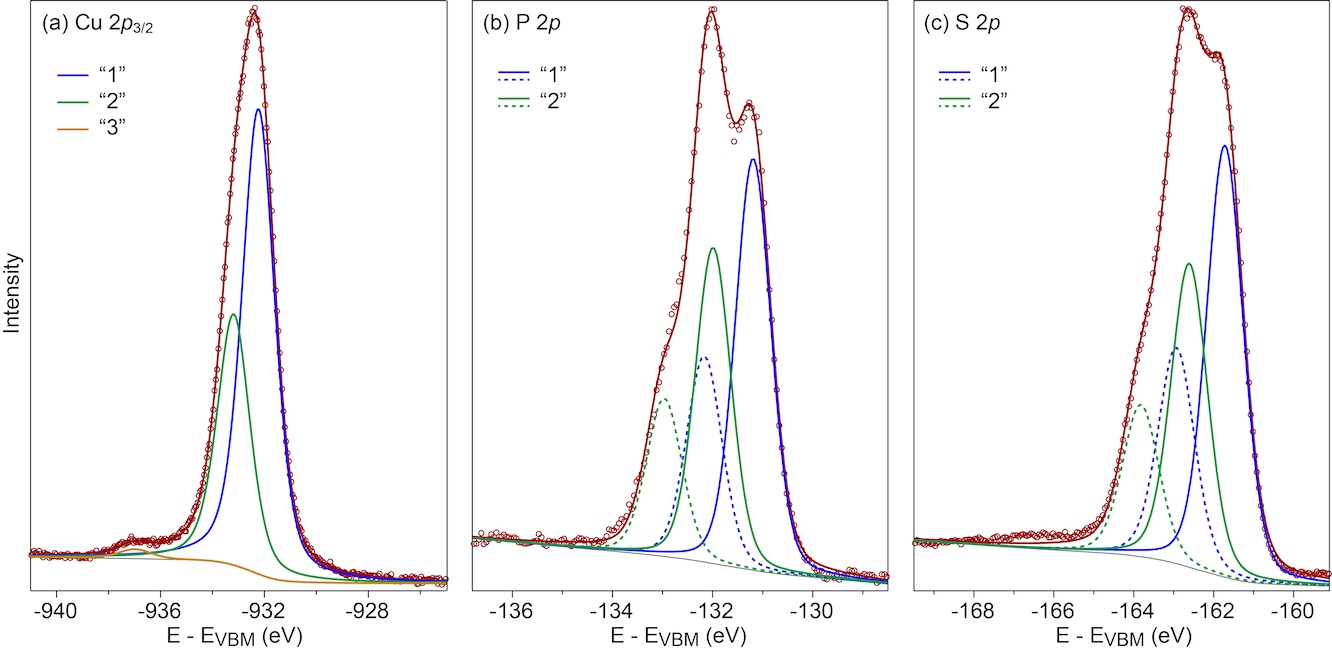}\\
Fig.\,S9. Results of fit procedure for core-level XPS spectra of CuCrP$_2$S$_6$ bulk crystal collected in the laboratory conditions: (a) Cu\,$2p_{3/2}$, (b) P\,$2p$, (c) S\,$2p$.
\end{figure}

\clearpage
\begin{figure}
\includegraphics[width=0.75\textwidth]{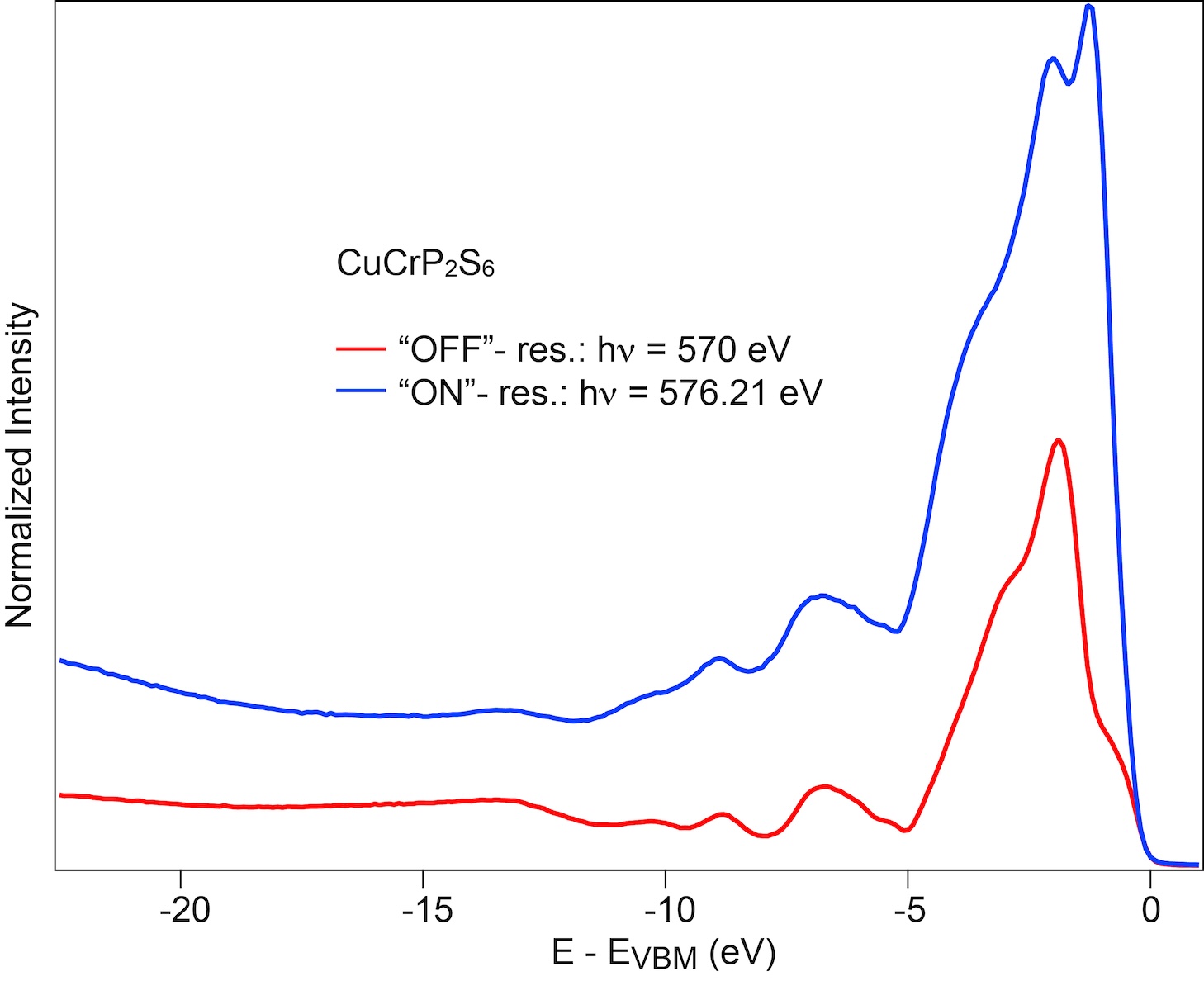}\\
Fig.\,S10. ``ON'' and ``OFF'' resonance spectra measured for CuCrP$_2$S$_6$ in a wide energy range at the Cr\,$L_{3}$ absorption edge. Photon energies are marked in the plot.
\end{figure}

\clearpage
\begin{figure}
\includegraphics[width=\textwidth]{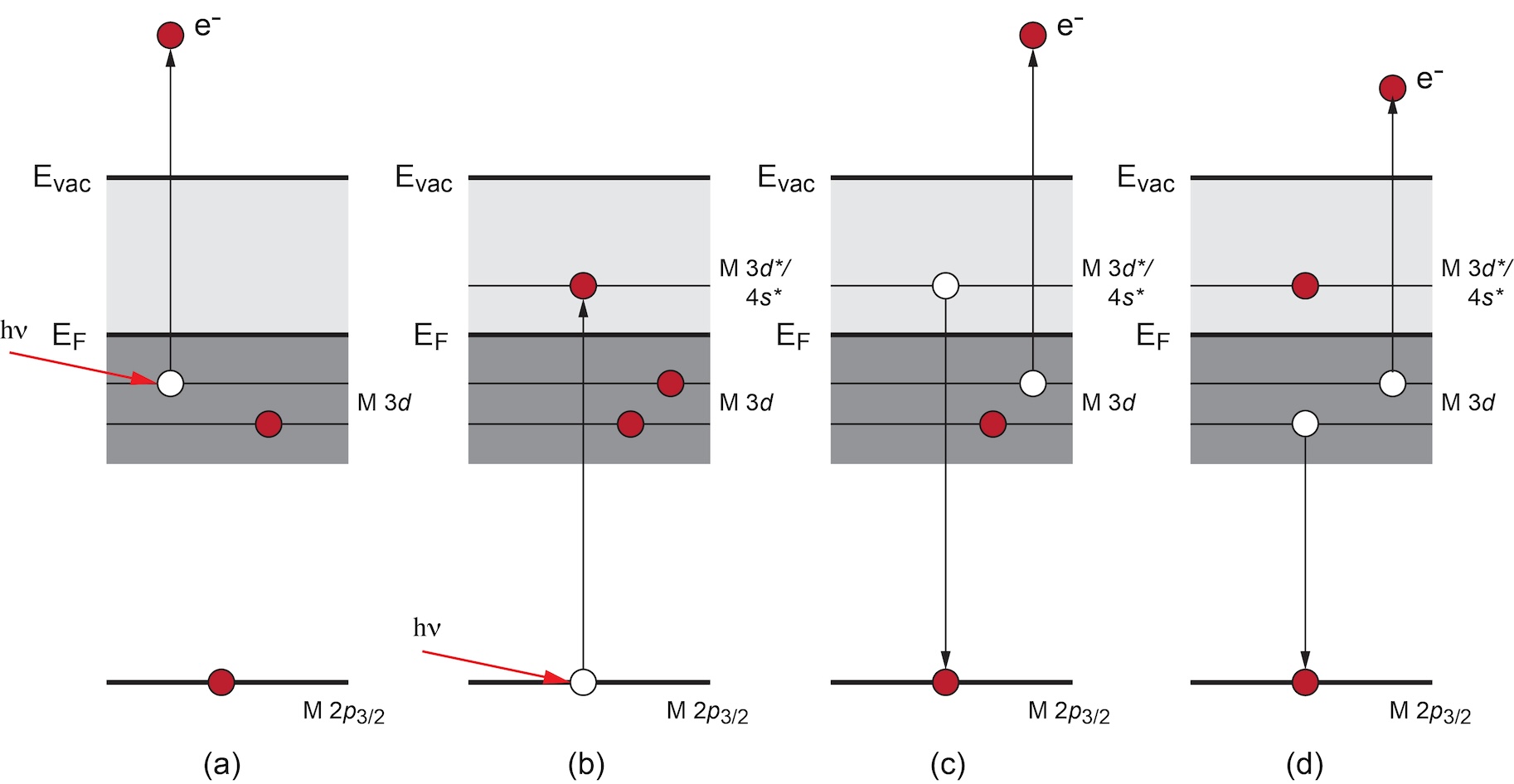}\\
Fig.\,S11. Schematic representation for processes during resonant photoemission: (a) direct photoemission channel; (b) core electron excitation process from $2p$ core level on unoccupied $3d^*/4s^*$ states above the Fermi level followed by (c) participator-Auger or (d) spectator-Auger electron decay processes of a resonant excitation.
\end{figure}

\clearpage
\begin{figure}
\includegraphics[width=0.75\textwidth]{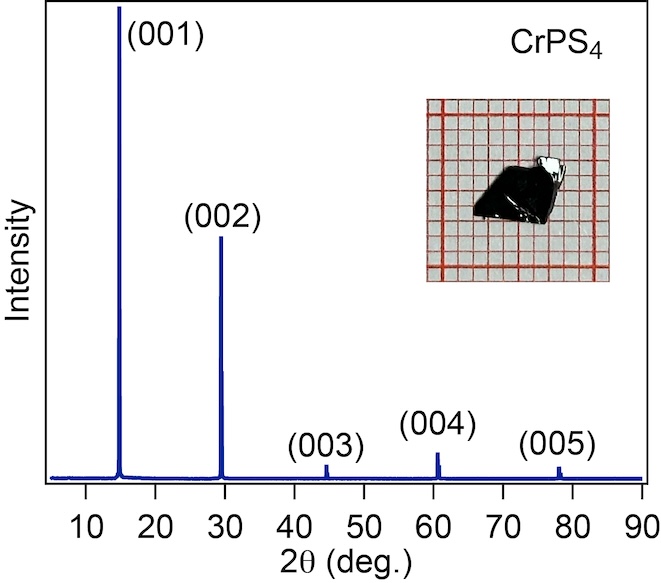}\\
Fig.\,S12. XRD diffraction spots for CrPS$_4$ bulk crystal. Inset shows the photo of crystal.
\end{figure}

\clearpage
\begin{figure}
\includegraphics[width=0.75\textwidth]{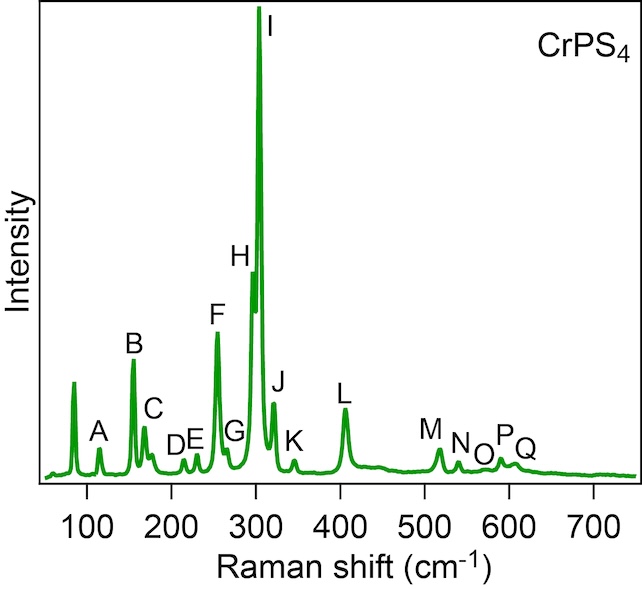}\\
Fig.\,S13. Representative Raman spectrum of bulk CrPS$_4$ crystal. Peak notations are taken from J. Lee \textit{et al.} Structural and Optical Properties of Single- and Few-Layer Magnetic Semiconductor CrPS$_4$. ACS Nano \textbf{11}, 10935 (2017).
\end{figure}

\clearpage
\begin{figure}
\includegraphics[width=0.75\textwidth]{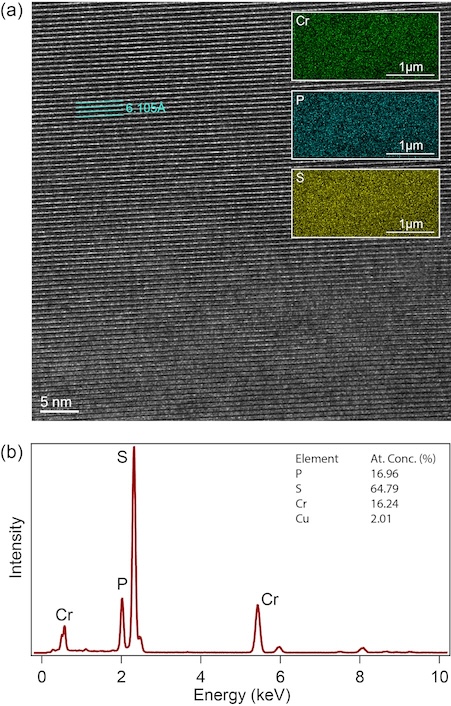}\\
Fig.\,S14. (a) TEM image of the CrPS$_4$ bulk crystal obtained in the same CVT synthesis. (b) The respective EDX spectrum of CrPS$_4$.
\end{figure}

\end{document}